\documentclass{aa} 
\usepackage[utf8]{inputenc}
\usepackage{graphicx}
\usepackage{amsmath,xcolor}

\begin{document}



\title{A new approach to spectroscopic phase curves}
\subtitle{The emission spectrum of WASP-12b observed in quadrature with HST/WFC3}
\author{J. Arcangeli \inst{\ref{inst1}}, J.-M. Désert \inst{\ref{inst1}}, V. Parmentier \inst{\ref{inst2}}, S.-M. Tsai\inst{\ref{inst2}}, K. B. Stevenson\inst{\ref{inst3}}}
\date{Month Year}
\authorrunning{Arcangeli et al.}
\titlerunning{Partial phase curve of WASP-12b with HST}

\institute{Anton Pannekoek Institute for Astronomy, University of Amsterdam, Science Park 904, 1098 XH Amsterdam, The Netherlands \label{inst1} \and Atmospheric, Ocean, and Planetary Physics, Clarendon Laboratory, Department of Physics, University of Oxford, Oxford, OX1 3PU, UK \label{inst2} \and JHU Applied Physics Laboratory, 11100 Johns Hopkins Rd, Laurel, MD 20723, USA \label{inst3}}

\abstract{
We analyse emission spectra of WASP-12b from a partial phase curve observed over three epochs with the Hubble Space Telescope, covering eclipse, quadrature, and transit, respectively. As the 1.1-day period phase curve was only partially covered over three epochs, traditional methods to extract the planet flux and instrument systematic errors cannot recover the thermal emission away from the secondary eclipse. To analyse this partial phase curve, we introduce a new method, which corrects for the wavelength-independent component of the systematic errors. Our new method removes the achromatic instrument and stellar variability, and uses the measured stellar spectrum in eclipse to then retrieve a relative planetary spectrum in wavelength at each phase.
We are able to extract the emission spectrum of an exoplanet at quadrature outside of a phase curve for the first time; we recover the quadrature spectrum of WASP-12b up to an additive constant.
The dayside emission spectrum is extracted in a similar manner, and in both cases we are able to estimate the brightness temperature, albeit at a greatly reduced precision, because our method removes the absolute level of the spectra, and therefore relies on fitting the slope of the emission spectrum instead of its amplitude.
We estimate the brightness temperature from the dayside (T$_{day}$=3186$\pm$677 K) and from the quadrature spectrum (T$_{quad}$=2124$\pm$417 K) and combine them to constrain the energy budget of the planet. We compare our extracted relative spectra to global circulation models of this planet, which are generally found to be a good match. However, we do see tentative evidence of a steeper spectral slope in the measured dayside spectrum compared to our models. We find that we cannot match this increased slope by increasing optical opacities in our models.
We also find that this spectral slope is unlikely to be explained by a non-equilibrium water abundance, as water advected from the nightside is quickly dissociated on the dayside.
\par

We present our technique for analysing partial or full phase curves from HST/WFC3 using common mode methods.
Importantly, and unlike previous phase curve analyses, this technique does not assume a functional form for the planet's emission with phase and does not require a full-orbit phase curve. The success of this technique relies upon stable pointing of the telescope between visits, with less than 0.1 pixels drift for example. This technique becomes powerful in the study of new regimes in exoplanetary systems such as for longer period planets, and is ideally suited for future observations with JWST and ARIEL.
}

\maketitle


\section{Introduction}

Measuring the thermal emission of planets provides key insights into the nature of their atmospheres, from their temperature structures to their compositions. The daysides of close-in transiting planets are the perfect examples of this because of the large signals emitted from their hot daysides and because an eclipse establishes a baseline from which the emission of the planet can be measured \citep{Charbonneau2005, Deming2006}. The natural extensions to dayside emission measurements are phase curves, where again the in-eclipse flux acts as a baseline from which thermal emission can be measured throughout the whole orbit of the  planet, thus probing the longitudinal temperature structure and chemistry \citep{Knutson2007, Borucki2009, Snellen2009}. The Hubble Space Telescope Wide Field Camera 3 (HST/WFC3) has allowed the expansion from photometric phase curves to spectroscopic phase curves in order to measure the phase-resolved spectra of hot planets on short periods \citep{Stevenson2014c, Kreidberg2018, Arcangeli2019}.

Phase-curve observations are time consuming because they rely on continuous monitoring  of the reflected or emitted light from an exoplanet during a full orbit around its host star. Due to their expensive nature and some technical limitations, phase curves have been limited to planets on short ($\sim$1 day) periods. In principle, measurements of exoplanet spectra at various longitudinal phases without obtaining a complete phase curve, so-called partial phase curves, are interesting for many scientific purposes. For example, partial phase curves are a unique opportunity to study the circulation on longer period planets, or can  simply be used to reduce the time required to observe planets on short periods. However, using current techniques at low spectral resolution, reflected or emitted light can only be measured when a baseline eclipse is present in each visit. Specifically, measurements away from the dayside, such as at quadrature, or of the nightside are inaccessible outside of full phase curves. Therefore, a new approach to removal of  systematic errors is required to observe partial phase curves and longer period planets in the future.

A common problem to phase curve observations is that instrument systematic errors can operate on the same timescale as the full orbit of a close in planet, typically one to two days in period \citep{Stevenson2014c}. For HST, while short-term systematic errors are accurately fitted by charge trap models \citep{Zhou2017}, the long-term systematic errors are thought to originate from changes in the thermal properties and optical path of the telescope, and are not well predicted. Furthermore, different models of the long-term systematic errors and thermal emission can lead to differences in the inferred atmospheric properties \citep{Kreidberg2018, Arcangeli2019}. Traditionally, these systematic corrections are handled using continuous monitoring of the planet as it orbits its parent star from eclipse to eclipse. This is done assuming the planet is tidally locked and that we therefore return to the  same hemisphere of the planet when observing consecutive eclipses. Recent work has highlighted methods for dealing with these systematic errors, such as observing multiple phase curves over different epochs to help characterise them \citep{Stevenson2014c, Kreidberg2018} or extending the phase curve to re-establish a baseline eclipse \citep{Arcangeli2019}. However, both these approaches require additional observing time, and limit the targets for which this technique can be applied to those with very short orbital periods.

Here, we present a new approach to phase-curve extraction that relies on common-mode methods and allows us to extract the emission from WASP-12b at different phases, including quadrature. Our new method removes the achromatic instrument and stellar variability, and uses the measured stellar spectrum in eclipse to then retrieve a relative planetary spectrum in wavelength at each phase. We apply our new method to the perfect test case for these issues: the partial phase curve of WASP-12b observed with HST/WFC3 at three different epochs over five days in 2011. Of these data, the dayside emission spectrum and the transmission spectrum have been published \citep{Swain2013}, but the emission at quadrature remains inaccessible to current methods. 

Interestingly, WASP-12b is a member of an emerging class of exoplanets referred to as the ultra-hot Jupiters (UHJs), which are distinguished by their extremely hot dayside temperatures resulting in dissociation- and ionisation-dominated chemistry \citep{Bell2017, Arcangeli2018, Kitzmann2018, Kreidberg2018, Mansfield2018, Parmentier2018, Evans2019, Baxter2020}. This makes the emission at different planetary phases extremely valuable, as it can be used to constrain the climate and atmospheric dynamics of the planet, and to probe the chemistry of the  planet beyond the muted features in the dayside spectrum.

We outline the data and our method in Section~\ref{Sec:Data} and explore the accuracy of the technique. Our results for WASP-12b are described in Section~\ref{Sec:Results}, along with a comparison to 3D forward models. We discuss the implications of our results and possible future applications in Section~\ref{Sec:Discussion}, and conclude in Section~\ref{Sec:Conc}.

\section{Data and Methods}
\label{Sec:Data}

\subsection{HST data}
\begin{figure*}
\includegraphics[scale=0.6]{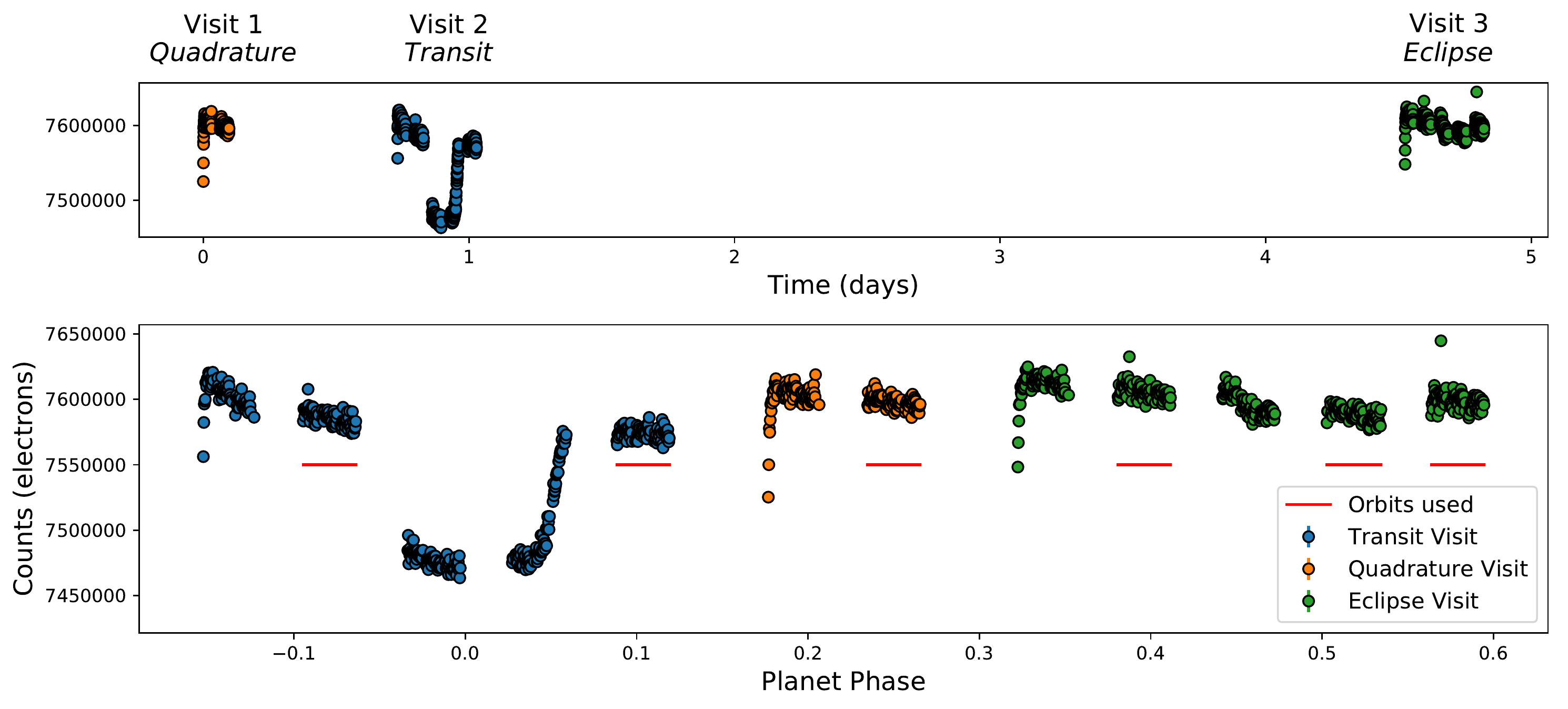}
\caption{Top: Raw phase curve plotted in time. Bottom: Raw phase curve phase folded to the period of the planet. The orbits used in the analysis are marked by a red line.}
\label{fig:ph_pc}
\end{figure*}

Our data set is the partial phase curve of WASP-12b from April 2012 (Program 12230, PI M.R. Swain, \citealt{Swain2013}).
These data consist of three visits: five orbits taken around transit, five orbits taken around secondary eclipse, and two orbits taken in quadrature (between transit and secondary eclipse). The raw data are shown in Figure~\ref{fig:ph_pc} and described in Table~\ref{Tab:Obs}. Together these visits constitute coverage of just over half of the phase curve of WASP-12b, with each visit taken within the same five-day period. The transmission spectrum and eclipse were published in \citet{Swain2013}, \citet{Mandell2013}, and \citet{Stevenson2014a, Stevenson2014b} while the visit taken in quadrature is unpublished. All visits were taken in staring mode with the same instrument setup, including a fixed orientation of the telescope.

\begin{table}[ht]
\begin{center}
\begin{tabular}{ | c | c | c |}
\hline
Start time & Observation & Planet phase \\
\hline 
2011-04-11 07:15 & Quadrature & 0.18 to 0.27 \\
2011-04-12 00:48 & Transit & -0.15 to 0.12 \\
2011-04-15 19:50 & Eclipse & 0.32 to 0.60 \\
\hline
\end{tabular}
\end{center}
\caption{Table of observed visits and corresponding planet phases.}
\label{Tab:Obs}
\end{table}

\subsection{Method}
\label{Sec:Method}

We use a two-step correction to extract the planet signal. First we use a common-mode correction to remove instrument systematic
errors, and second we compare the extracted spectra at each orbit to the stellar spectrum measured in eclipse in order to measure the planet spectra. For the first step, we divide each spectroscopic light curve by the white-light curve (integrated over the full wavelength range), removing any common-mode systematic
errors. In the second step, we divide the spectra of each orbit by a reference spectrum of the star, taken during secondary eclipse and processed in the same way. This produces a relative spectrum of the planet at each phase up to an additive constant. We refer to these spectra as relative spectra, referring to the fact that they are relative in wavelength, as their absolute flux levels are removed by the common-mode correction. Spectra are not extracted for the first orbit of each visit, as they are seen to exhibit stronger systematic
errors than subsequent orbits.

Below we illustrate the applied method. The raw data $L_{\lambda, t}$ and $w_{t}$, the spectroscopic light curves and white light curve respectively, can be expressed by the formulae below.

\begin{align}
& L_{\lambda, t} = S_{\lambda, t} * [ F^{star}_{\lambda, t} + F^{planet}_{\lambda, t} ] \nonumber \\
& w_{t} = s_{t} * [ f^{star}_t + f^{planet}_{t} ].
\label{eq:definitions}
\end{align}

Here, $S_{\lambda, t}$ and $s_t$ are the unknown systematic effects, and $F_{\lambda, t}$ and $f_t$ are the spectra and white-light flux output, respectively, of either the star or the planet. We assume that the star remains constant in time over the visit, with the exception of the in-transit data which we exclude, that is, $f^{star}_t=f^{star}$ and $F^{star}_{\lambda, t}=F^{star}_{\lambda}$.

One can therefore use the white-light curve to divide out the unknown time-dependent systematic component. We now assume that $S_{\lambda, t}=s_t*c_{\lambda}$, where $c_{\lambda}$ is a normalisation constant for each wavelength bin, which means that we assume that the systematic
errors in the white-light curve are representative of each wavelength bin, often referred to as a common mode. We separately test the validity of this assumption in Section~\ref{Sec:Rednoise}. Dividing Equations~\ref{eq:definitions} becomes:

\begin{align}
L'_{\lambda, t} = L_{\lambda, t} / w_t = c_{\lambda}*[ F^{s}_{\lambda}+ F^{p}_{\lambda, t} ] / [f^{s} + f^{p}_{t} ],
\label{eq:corrected_lc}
\end{align}

where $L'_{\lambda, t}$ are now the common-mode corrected light curves. Evaluating Equation~\ref{eq:corrected_lc} at $t=t_{eclipse}$ gives:

\begin{align}
L'_{\lambda, t=t_{ecl}} = c_{\lambda}*F^{s}_{\lambda} / f^s,
\label{eq:in_eclipse}
\end{align}

where $F^{s}_{\lambda}$ is the stellar spectrum, and $f^s$ is the white-light flux in eclipse.

Therefore, using the in-eclipse data we can measure the spectrum of the star multiplied by the remaining wavelength-dependent systematic
errors. This then allows us to finally obtain the emission spectrum of the planet, which is achieved by dividing the spectra from the corrected light curve ($L'_{\lambda, t}$) at each time by the in-eclipse stellar spectrum:

\begin{align}
L'_{\lambda, t} / L'_{\lambda, ecl} & =  [ F^{s}_{\lambda}+ F^{p}_{\lambda, t} ] / F^{s}_{\lambda} * f^s / [f^{s} + f^{p}_{t} ] \nonumber \\
& = [ 1 + F^{p}_{\lambda, t} / F^{s}_{\lambda} ]  * [1 + f^{p}_{t}/f^s ]^{-1} \nonumber \\
& = 1 - f^{p}_{t}/f^s + F^{p}_{\lambda, t} / F^{s}_{\lambda} + \mathcal{O}((F^p/F^s)^2) \nonumber \\
& \sim q(t) + F^{p}_{\lambda, t} / F^{s}_{\lambda.}
\label{eq:final}
\end{align}

The parameter $q(t)=1-f^{p}_{t}/f^s$ is some time-dependent residual (independent of wavelength) that remains after dividing out the wavelength-dependent component of the systematic
errors, and can be removed by re-normalising the extracted spectra at each orbit. This last step of ignoring higher order terms only holds for systems where the variability of the star is sufficiently small, which we discuss in Section~\ref{Sec:Star}. 

\subsection{Comparison to classical eclipse spectroscopy}

When applied to the secondary eclipse data of WASP-12b, our method is substantially different from the classical method, which is that used in \citet{Stevenson2014a}. As we do not fit for the depth of the eclipse, but rather use the out-of-eclipse data to measure the spectrum of the planet, our method is more akin to current spectroscopic phase curve extractions \citep{Stevenson2017, Kreidberg2018, Arcangeli2019}. There are still key differences between our approach and a typical phase-curve extraction, as we do not assume a functional form for the planet's emission with phase, or for the systematic
errors; nor do we recover the absolute spectrum of the planet at each phase, only the relative spectrum in wavelength. This approach is chosen so as to extract the spectra at each phase in the same manner, from the dayside spectrum to the nightside and quadrature spectra, as in this case neither a classical eclipse depth measurement nor a phase-curve fit can be applied to the whole data set because the observations are spread over five days. Therefore, this approach should be considered as a complement to the existing extraction of the dayside emission and transmission spectra.

Our approach has consequences for how we compare to circulation models (such as those present in Section~\ref{Sec:models}). A typical dayside eclipse-depth measurement is normally compared to the magnitude of the phase curve at phase 0.5, as the eclipse depth is an average of the planet's emission across the dayside. In our approach, we extract the relative emission spectra before and after eclipse, and therefore compare to the circulation models at phases immediately before and after eclipse. This technique is then also applicable to every other phase of the planet's emission, with the exception of the in-transit data, as each phase can be directly compared to the same phase in the models.

\subsection{Pointing offsets}
\label{Sec:Offsets}
\begin{figure*}
\includegraphics[scale=0.65]{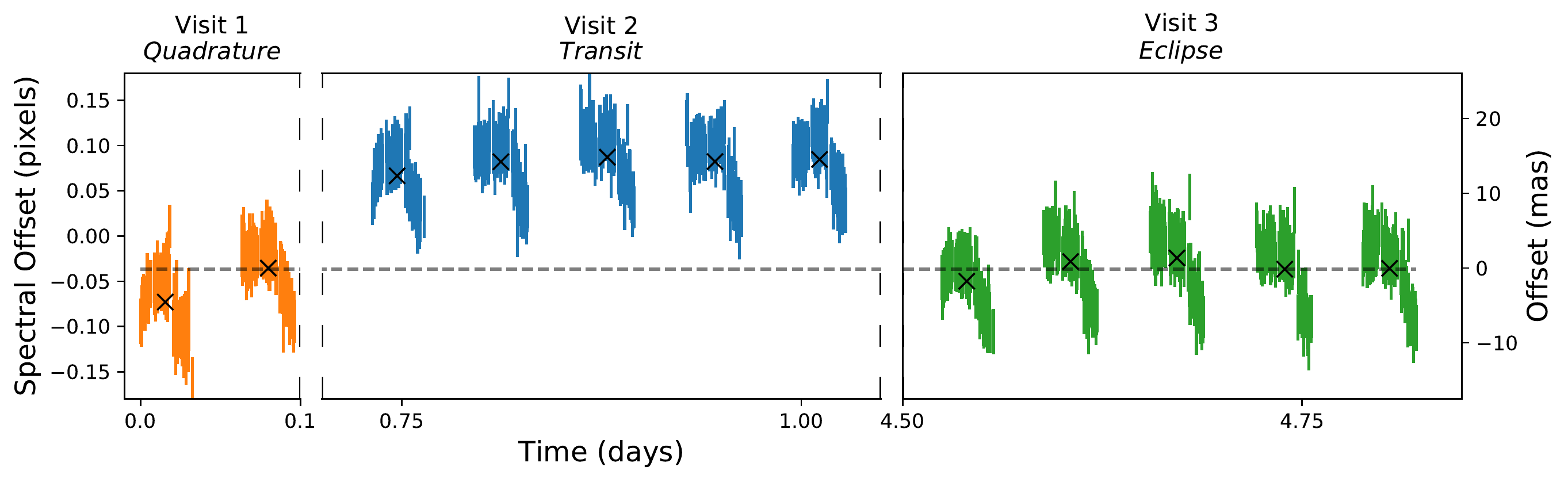}
\caption{Offset of spectrum position in the dispersion direction relative to average over all exposures as described in Section~\ref{Sec:Offsets}. The x-axis shows the time in days, with line breaks between different visits.} Coloured lines indicate the offset for each exposure with 1$\sigma$ uncertainties. Black crosses are the average offset for each orbit, with the horizontal dashed line set to the quadrature orbit offset for visual reference.
\label{fig:shifts}
\end{figure*}

Shifts of the spectra in wavelength can dominate the systematic
errors of light curves observed with HST/WFC3 \citep{Wakeford2016, Tsiaras2016, Tsiaras2019}. In addition, our technique relies on the stellar spectrum remaining constant between visits, which requires that the position of the spectrum on the detector also remains constant, or can be accurately reconstructed from a shifted spectrum. The IR WFC3 pixel flat-field is uncertain at the 1\% level, before considering any correlation between the flat-fielding of nearby pixels, and the spectrum observed is under-sampled even before binning by 6 pixels per bin as we do in our analysis (see \citealt{Deming2013}). Therefore, measuring and interpolating any shifted spectra back to a reference position on the detector is increasingly difficult for larger positional shifts. Typically, a drift of <15 mas  (0.11 pixels) in the x-position is optimal for exoplanet studies using the spatial scanning mode \citep{Stevenson2019}.

We first assess the possible effect of shifts on our spectra by measuring the position of the dispersed spectrum on the detector for each exposure. We take the flat-field-corrected raw images and perform a column-sum to produce a raw spectrum for each exposure. We then produce a reference spectrum from the average of all the raw spectra, and cross-correlate between this and each of the raw-exposure spectra to measure their positional shifts on the detector. The measured offsets are shown in Figure~\ref{fig:shifts} in pixels along the x-axis.

\begin{figure}
\includegraphics[scale=0.45]{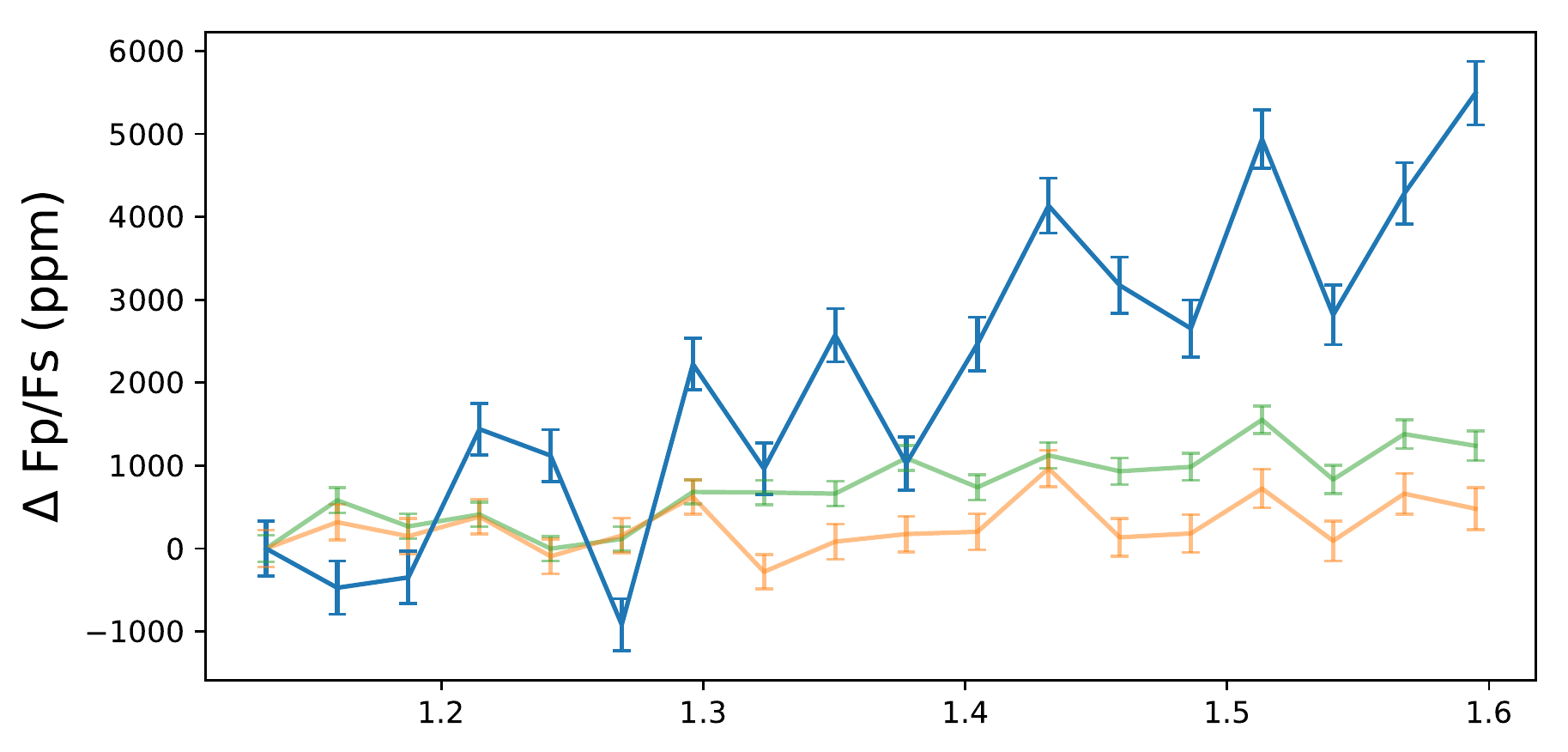}
\includegraphics[scale=0.45]{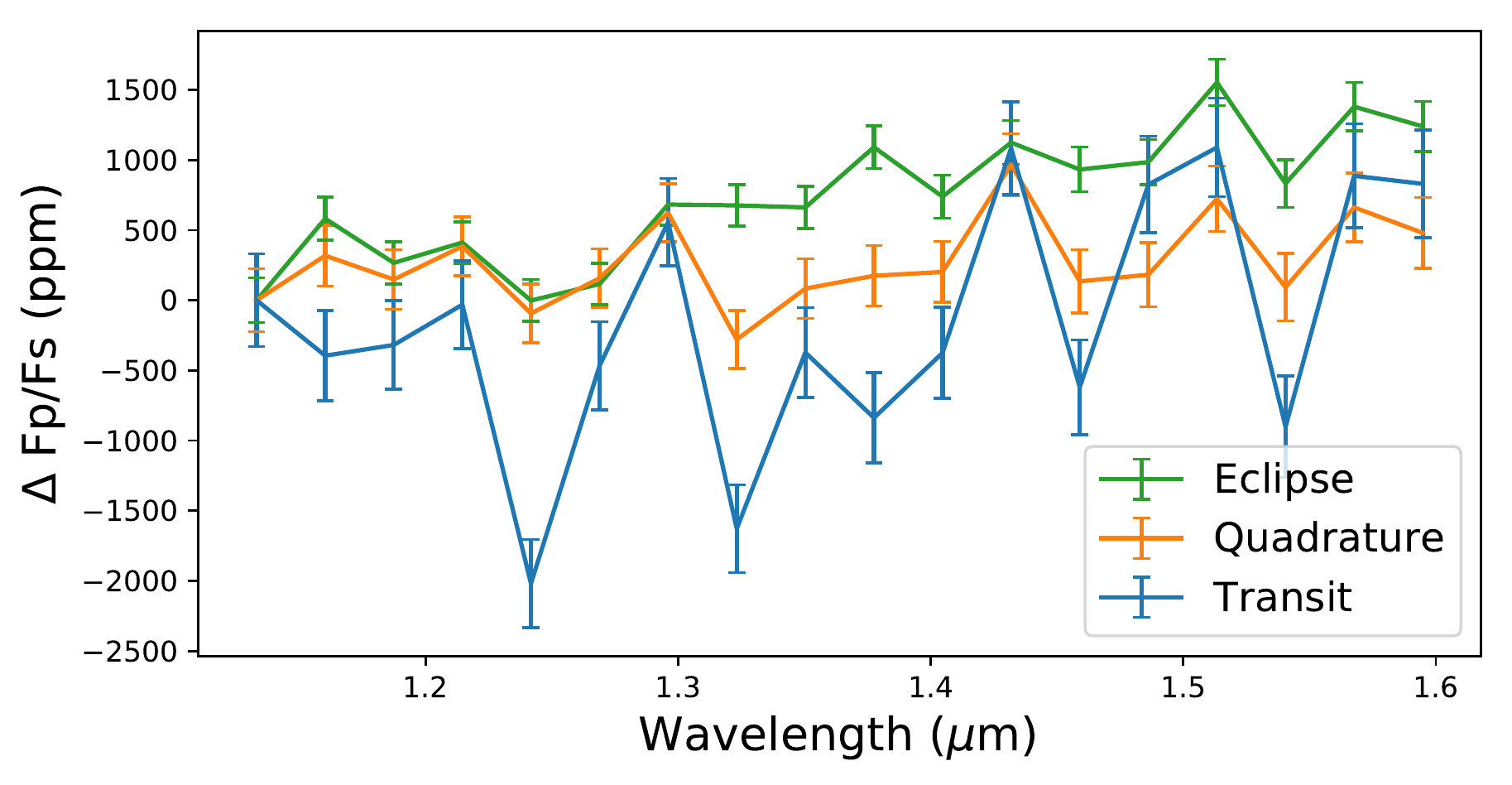}
\caption{Top: Common-mode corrected exoplanet spectra with propagated 1-$\sigma$ error bars as described in Section~\ref{Sec:Method}. The transit spectrum extraction in blue is referred to as the nominal extraction. Bottom: Same spectra, except for the nominal transit spectrum which has been re-interpolated by the measured positional shift (computed in Section~\ref{Sec:Offsets}). The systematic spectral slope is largely removed but residual noise is still clearly present in the spectrum in the form of variation between spectral bins with amplitude larger than the expected errors or planet signal.}
\label{fig:corrspecs}
\end{figure}

For the transit visit, we find that there is a significant offset in x-position relative to the other visits, that is, of greater than 0.1 pixels. If uncorrected, this offset produces a slope in the extracted spectra which we estimate to be on the order of 0.1\% (1000 ppm), which is due to the combination of the under-sampling of the spectrum and flat-field uncertainties (discussed further below). In principle, the transit spectra could be interpolated back to remove this slope; however, in practice the accuracy of this interpolation is limited, leading to an additional uncertainty on the transit spectrum. We re-interpolate the spectra obtained during the transit visit to the reference in-eclipse exposure, and plot the results in Figure~\ref{fig:corrspecs}. Here we see that the emission spectrum close to transit is dominated by large variations between bins above the expected noise level, and is not usable.

We can first estimate the effect of flat-field errors on our spectra by using the nominal flat-field uncertainty of 0.2\% per pixel \citep{Bushouse2008}. This corresponds to an uncertainty of 200 ppm on the pixel flux for a shift of 0.1 pixels.

\begin{figure}
\includegraphics[scale=0.5]{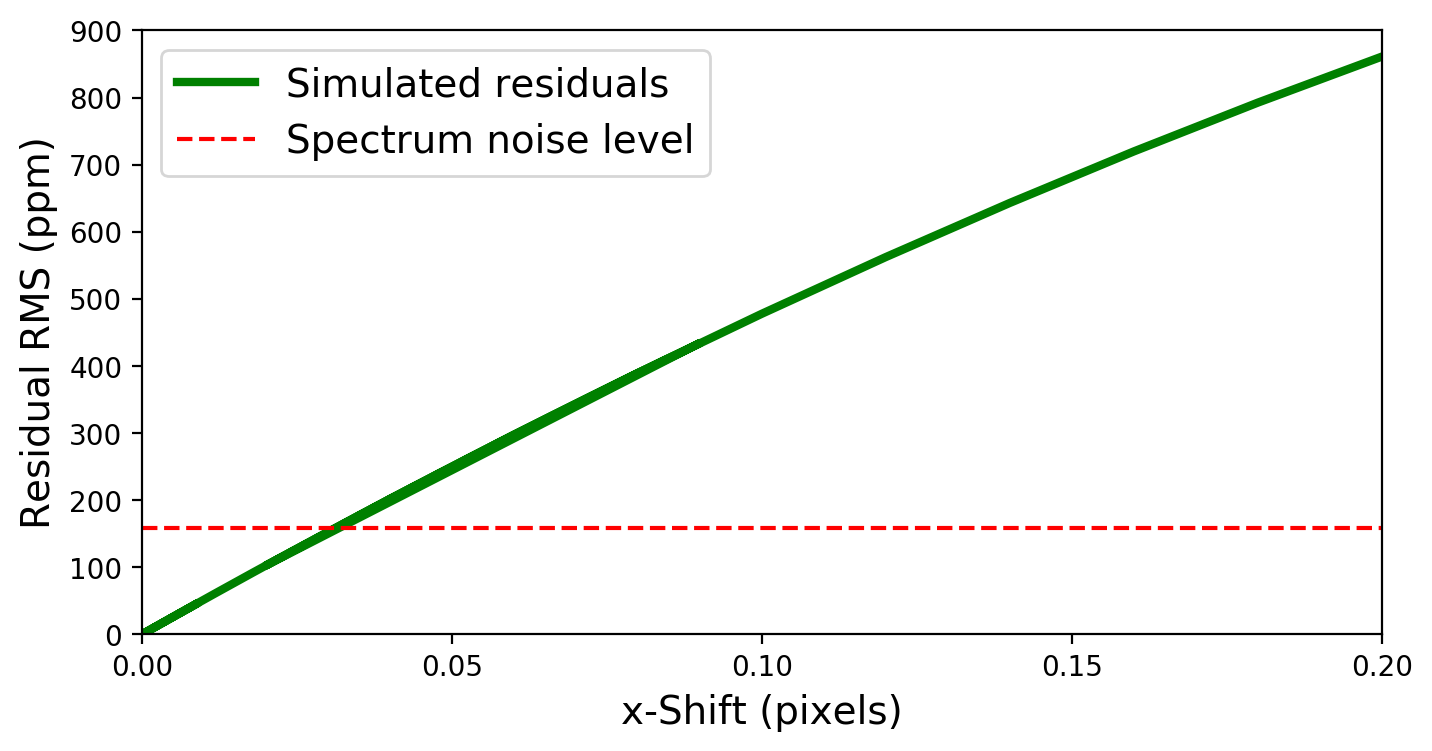}
\caption{Simulated residuals as caused  by spectral under-sampling  (green curve). The computation is made by a convolution of a stellar spectrum at the resolution of the data, which is subsequently interpolated for each positional shift. The dashed red line is the noise level of the extracted eclipse spectrum. This simulation does not include the propagation of flat-field uncertainties, which will act to further increase the residual noise.}
\label{fig:shiftsim}
\end{figure}

To illustrate the effect of under-sampling on our extracted spectra, we test its effect on a model stellar spectrum of WASP-12b. The model spectrum is taken from the Kurucz atlas of stellar models, using nominal parameters for the star WASP-12A \citep{Kurucz1993}. We produce stellar spectra at the resolution of the data for a range of shifts in spectral pixels using the pysynphot package \citep{pysynphot}. We then re-interpolate each shifted spectrum to a reference unshifted spectrum, take the ratio between the interpolated spectrum and the reference, and then compute the amplitude of the remaining residuals. This amplitude, shown in Figure~\ref{fig:shiftsim}, can be seen as the additional noise introduced by under-sampling of a shifted spectrum.  Figure~\ref{fig:shiftsim} includes the residual noise level of the eclipse spectrum (158 ppm), shown as a line in red. Here we can see that under-sampling errors dominate the spectral noise above shifts of 0.03 pixels.

We also measure the positional shift of the spectrum on the detector between the quadrature orbit and the in-eclipse orbit to be negligible at a precision of  0.0011
pixels, as seen in Figure~\ref{fig:shifts}. This translates to an uncertainty of 11 ppm per pixel from the flat-field correction, which is well within the errors on the quadrature spectrum. We can therefore conclude that the systematic
errors 
induced by 
positional-drift should be negligible for the quadrature spectrum.


\subsection{Residual noise}
\label{Sec:Rednoise}
We estimate the quality of our systematic correction by comparing the  in-eclipse orbit corrected for systematic
error to a flat line at each wavelength. By measuring the scatter of the corrected in-eclipse orbits compared to our expectation from photon noise, we can place an upper limit on the residuals caused by any wavelength-dependent components of the systematic
errors over a short baseline.

\begin{figure}[h]
\includegraphics[scale=0.47]{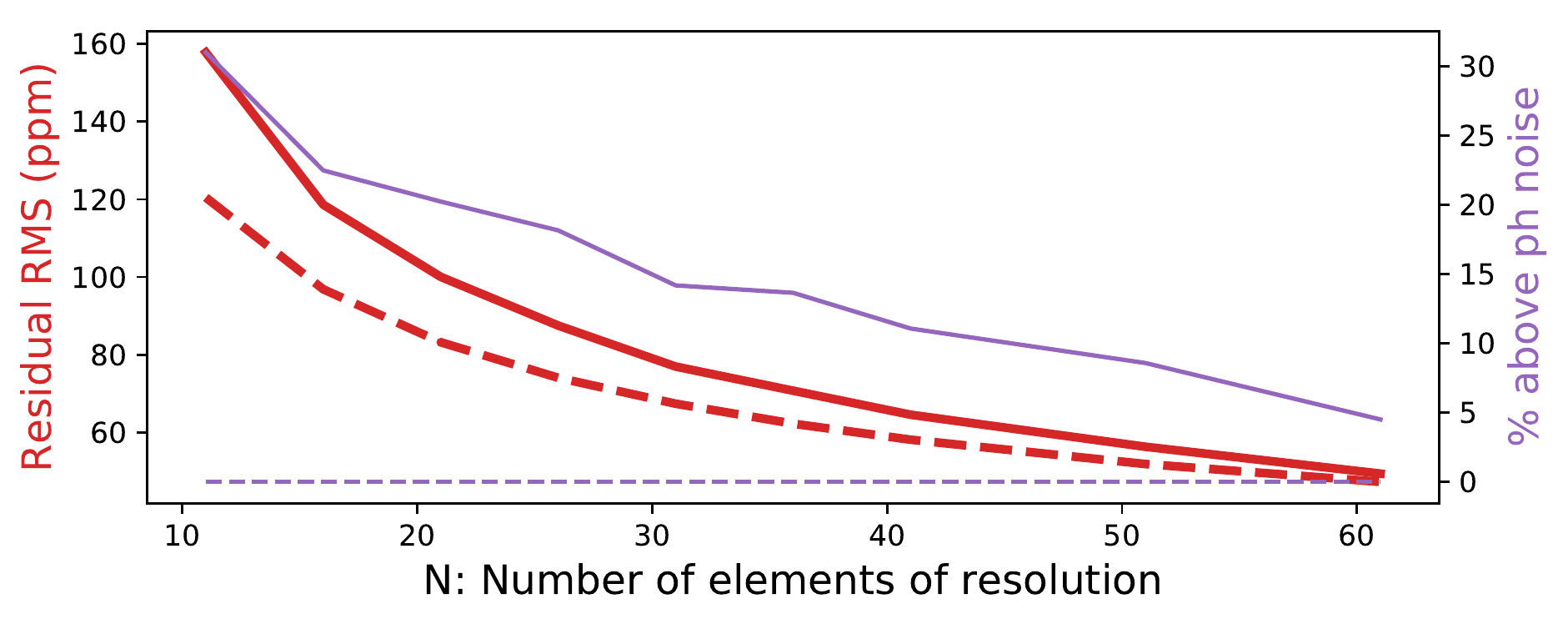}
\caption{Measured scatter of the residuals of the light curve taken from the in-eclipse orbit and normalised by the white-light curve for various sizes of wavelength bin (decreasing number of wavelength bins towards the right). The scatter represents the average rms of the residuals from the light curve across the wavelength bins for each choice of binning. The left y-axis shows the residual rms,  plotted as the thick red lines, while the right y-axis shows the scatter as a percentage of the photon noise limit, plotted as the thin purple lines. Dotted lines show the expected photon noise limit in each case.}
\label{fig:rednoise}
\end{figure}

We measure the scatter between the exposures at each wavelength for an increasing number of wavelength bins, shown in Figure~\ref{fig:rednoise}. For each wavelength bin, we divide by a reference white-light curve from the sum of the other wavelength bins and examine the residuals. We then measure the scatter of the residuals at each wavelength, and report the average scatter for each wavelength bin size. We find that the average scatter reaches close to photon noise precision, around 20\% above photon noise, for 15 or more wavelength bins during the eclipse. We choose to bin by 6 pixels, corresponding to 18 total wavelength bins over the full spectrum, to maximise the signal-to-noise ratio (S/N) of our final spectra, while maintaining a precision of better than 20\% above photon noise.

\subsection{Dilution correction}

The WASP-12 system has been found to include a faint companion binary system (WASP-12BC) to the main star (WASP-12A, \citealt{Bergfors2011, Bergfors2013, Crossfield2012a}).
We correct our extracted planet spectra for the dilution by the companion stars following the methods outlined in \citet{Stevenson2014a,Stevenson2014b}. This correction is made by measuring the ratio of the stellar spectra which in turn is done by simulating a range of models, accounting for their offset in position on the WFC3 detector. We then divide their contribution from each of our spectra, which can be found in Table~\ref{Tab:Dilution} and is roughly equivalent to a slope of about 100ppm across the whole wavelength range.

\subsection{Stellar variability}
\label{Sec:Star}

\begin{figure}[h]
\includegraphics[scale=0.4]{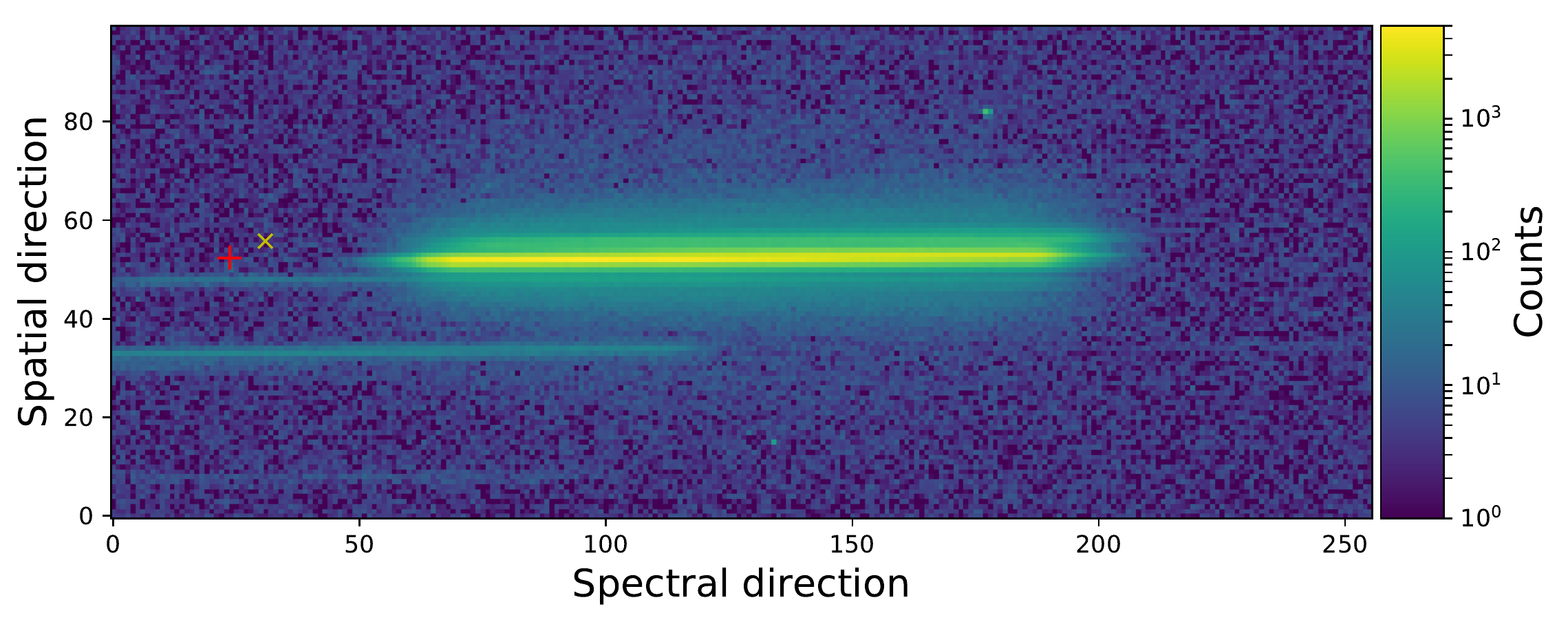}
\caption{Cutout 2D image of a single representative exposure. The bright spectrum in the centre is that of WASP-12A, and directly superimposed is the combined spectrum of WASP-12BC, seen as the faint trace above the main trace. Below these two are spectra of faint background stars. The locations where the non-dispersed direct images would fall are marked as crosses, red for WASP-12A and yellow for WASP-12BC.}
\label{fig:field_image}
\end{figure}

We asses the impact of any variability in the spectra of either WASP-12A or its companion stars WASP-12BC on our final planet spectra. We use an extraction window that contains the flux from WASP-12BC and WASP-12A, as the spectra bleed into each other on the detector (see Figure~\ref{fig:field_image}); this is why the variability of the companions needs to be considered. While any photometric variability in the stars is removed by the common mode correction, changes to the stellar spectra are not and could be left as residuals in our final planet spectra.

\begin{figure}[h]
\includegraphics[scale=0.48]{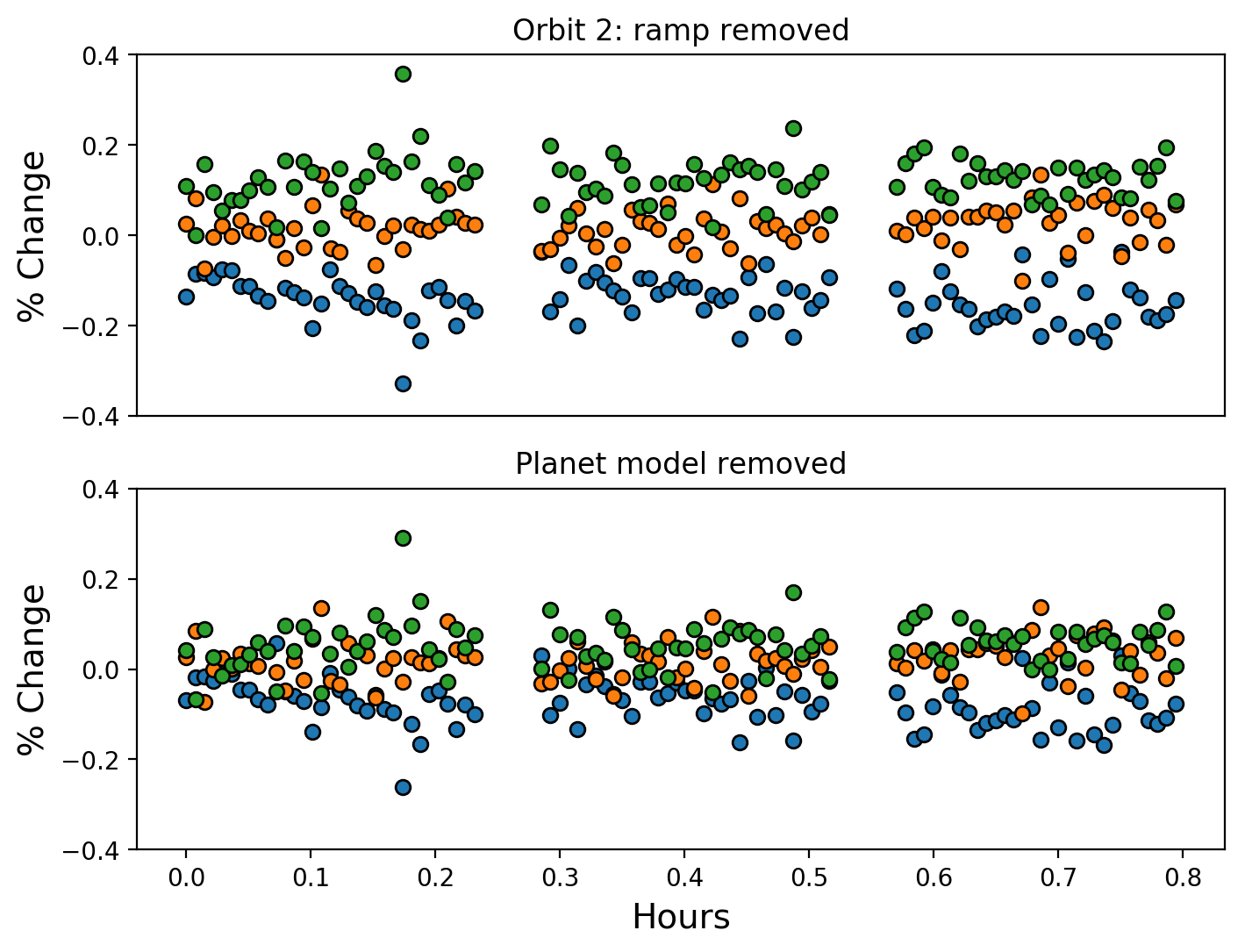}
\caption{Brightness of exposures taken during the second orbit of each visit, coloured by the corresponding visit and measured relative to an average across all three visits, discussed further in Section~\ref{Sec:Star}. Green points are from the transit visit, orange from the quadrature visit, and blue from the eclipse visit. The exposures are plotted in hours from the beginning of the orbit. Top: Exposures from the second orbits as a percentage change from the average second orbit across all visits. Bottom: Same exposures with the predicted planetary contribution removed.}
\label{fig:starvar}
\end{figure}

We estimate the variability of the WASP-12 system over the course of our observations by comparing the second orbit of each visit. In Figure~\ref{fig:starvar}, we plot the second orbit of each visit minus the average of the second orbits across all visits. This subtraction removes the orbit-ramp that is constant between orbits. The spread between the visits at this stage is approximately 0.13\% (top panel of Figure~\ref{fig:starvar}); however, a large fraction of this variability is expected to be due to the changing emission of the planet itself. We remove the predicted contribution from the thermal emission of WASP-12b, with the resulting stellar contribution shown in Figure~\ref{fig:starvar}, based on our nominal circulation model presented in Section~\ref{Sec:models}. The remaining spread between visits is 0.08\%, which is close to the photon noise limit, with the remaining variation consisting of a combination of changes in systematic
errors and stellar variability between visits. AS the planet model is uncertain, we adopt a conservative upper limit of 0.1\% on the variability of the WASP-12 stars within our observing window.

We estimate the impact that 0.1\% variability in the brightness of the WASP-12 system would have on our planet spectra. Assuming starspot temperatures of 300K cooler than the stellar surface of WASP-12A \citep{Kreidberg2015}, a 0.1\% change in brightness could be achieved by a 0.6\% change in spot covering fraction between visits. This would result in a change to the slope of the spectrum of the system of 50-100 ppm. Equivalently, a 1\% change in the brightness of the companions WASP-12BC would result in a 0.1\% change in the brightness of the system, leading to a similar slope in the extracted spectra of 50-100 ppm between visits. An additional slope of this magnitude or lower would be well within the uncertainties on the estimated blackbody temperatures, within 150-200 K, and would not change the results of this work.

\section{Results}
\label{Sec:Results}

\begin{figure}
\includegraphics[scale=0.48]{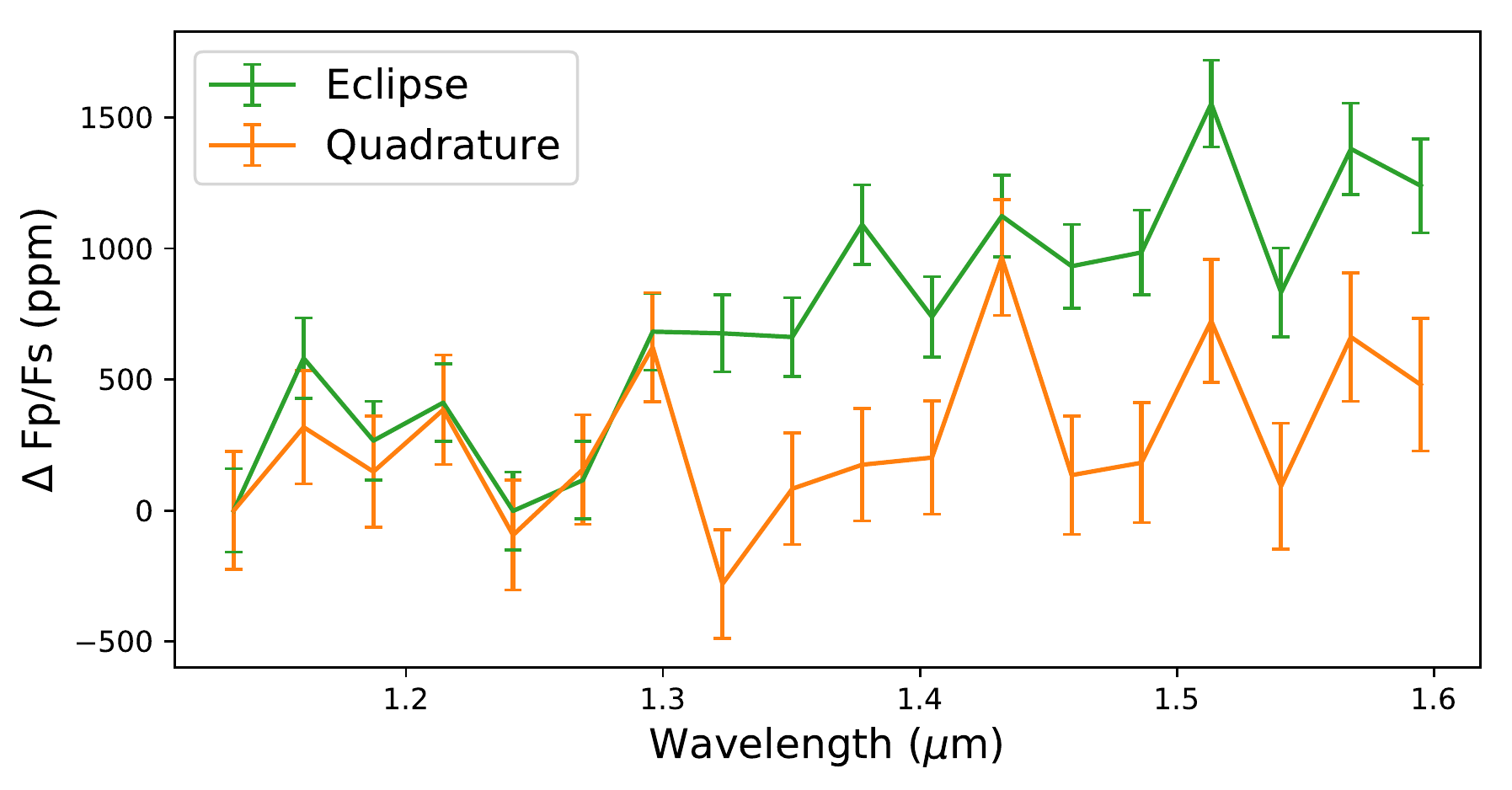}
\caption{Extracted emission spectra from quadrature and eclipse for WASP-12b, normalised to the first spectral bin.}
\label{fig:specs}
\end{figure}

\subsection{Extracted emission spectra}

We apply our new method of extracting the longitudinally resolved spectra from spectroscopic phase curves to the partial phase curve of WASP-12b. We extract the dayside spectrum before and after eclipse, as well as, for the first time, the quadrature spectrum of the planet (see Figure~\ref{fig:specs} and Table~\ref{Tab:Specs}). We find that the emission spectra from the dayside before and after eclipse are consistent, and therefore we combine them into one spectrum that we refer to as the eclipse or dayside spectrum, although we note this spectrum is not derived using classical eclipse spectroscopy. We first fit the spectra with black-body curves to estimate their corresponding temperatures from their spectral slopes, including an additive constant as a free parameter in each fit to account for the unknown absolute level of the spectra. We scale the uncertainty on the final blackbody temperatures by the square root of the chi-squared value to account for the quality of the fits. The best-fit dayside black body is at 3186$\pm$677 K at a reduced $\chi^2_r$ of 2.73, and the best-fit quadrature black-body is cooler at 2124$\pm$417 K with a $\chi^2_r$ of 1.77. Our extracted spectra show a decrease in slope between eclipse and quadrature, equivalent to a decrease in brightness temperature of 1062$\pm$795 K, which tentatively suggests weak redistribution.

\subsection{Comparison to models}
\label{Sec:models}
We compare our data to global circulation models (GCMs) of the atmosphere of WASP-12b. The atmospheric circulation and thermal structure were simulated using SPARC/MITgcm \citep{Showman2009}; further details on the models can be seen in \citet{Parmentier2018}. We produced two self-consistent forward models, one at a solar composition and one with enhanced optical opacities (by a factor of two). The enhanced optical opacity model was chosen to explore the effect of larger temperature inversions at low pressures. We also post-process our solar composition model with different opacities and physics removed in order to test the effects of H-, dissociation, and molecular opacities on our model spectra.

\begin{figure}
\includegraphics[scale=0.5]{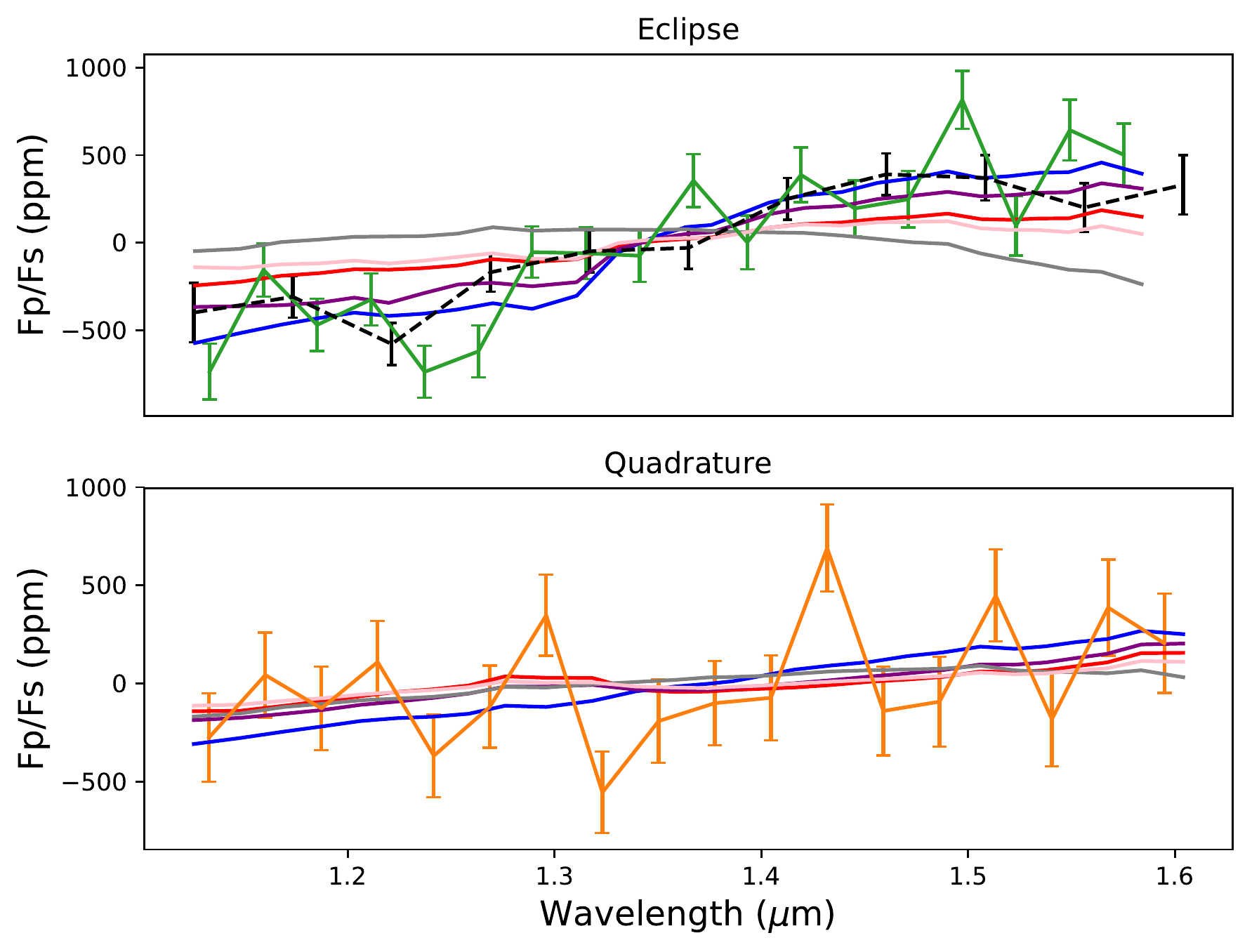}
\caption{Comparison of extracted exoplanet spectra to GCM results. All measured spectra and models are normalised by the average of the spectrum across the bandpass in order to allow a comparison to our relative spectra. Top: Data for the eclipse (in green) as well as the dayside spectrum extracted by \citeauthor{Stevenson2014b} (2014, in dashed black). Bottom: Extracted quadrature spectrum (in orange). For both panels the models are shown for solar composition (red), no-dissociation (blue), no H- (purple), no molecular opacity (grey), and factor-two higher optical opacity (pink).}
\label{fig:gcm_spec}
\end{figure}

As our measured spectra are only known relative to a spectral bin, we choose to normalise the spectra from both the data and models by subtracting the average level, shown in Figure~\ref{fig:gcm_spec}. We find that there is relative agreement between the observed quadrature spectrum and each of the models, while the observed eclipse spectrum appears slightly steeper than the models. We discuss the possible source of this difference in Section~\ref{Sec:slope}. However, it should be noted that the circulation models provide an adequate match to our observations in both cases, given that the GCMs themselves are not fitted to the data. For instance, these circulation models do not explore a large enough range of parameters to reach a `fit quality' comparable to a best-fit model that would be generated from a retrieval framework.

\subsection{Comparison to dayside emission studies}

\begin{figure}[h]
\includegraphics[scale=0.5]{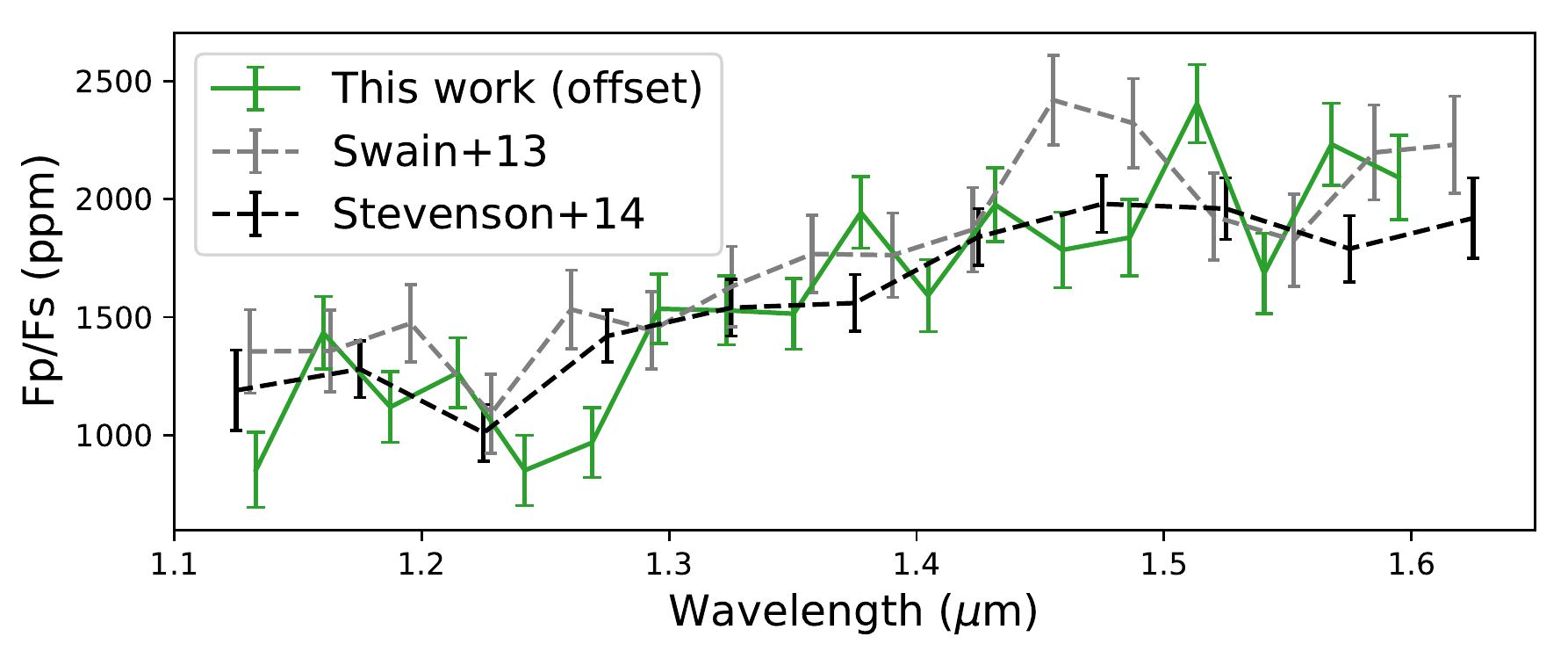}
\caption{Comparison of our dayside emission spectrum to previous those of previous studies, all extracted from the same data with different pipelines and techniques (see Section~\ref{Sec:Method}; \citealt{Stevenson2014b, Swain2013}). Our extracted relative spectrum has been offset to match the absolute level of \citet{Stevenson2014b}. All of the works are in agreement, with additional slight differences arising because of the choice of binning. }
\label{fig:eclipses}
\end{figure}

We can also compare our extracted dayside emission spectrum to previously published work using the same data \citep{Swain2013, Stevenson2014b}. The three dayside emission spectra are shown in Figure~\ref{fig:eclipses}. We can see that in general there is a good agreement between the spectra, with slight differences due to the choice of binning. This is notable because our approach is very different from that of these two latter studies. In classical secondary eclipse studies, such as those of \citet{Swain2013}, \citet{Stevenson2014b}, the dayside emission is measured from the difference between the out-of-eclipse orbits and the in-eclipse orbits. This is therefore an estimate of the emission from the full dayside at planetary phase 0.5. Our method instead measures the emission spectrum of the planet at phases $\sim$0.46 and 0.58 relative to a spectral bin. We find these spectra to be consistent, likely because they are dominated by the same hotspot, and therefore combine them into a single relative emission spectrum for the dayside.

The error bars on our fitted temperatures appear significantly larger than the uncertainties on the brightness temperatures previously extracted for WASP-12b, which were previously sufficiently small to be dominated by the uncertainty on the stellar temperature. The additional uncertainty present in our extracted temperatures originates from our method, which fits a black-body temperature to a relative spectrum and  must therefore fit for an additional parameter which is the absolute offset of the black body.

\begin{figure}
\includegraphics[scale=0.5]{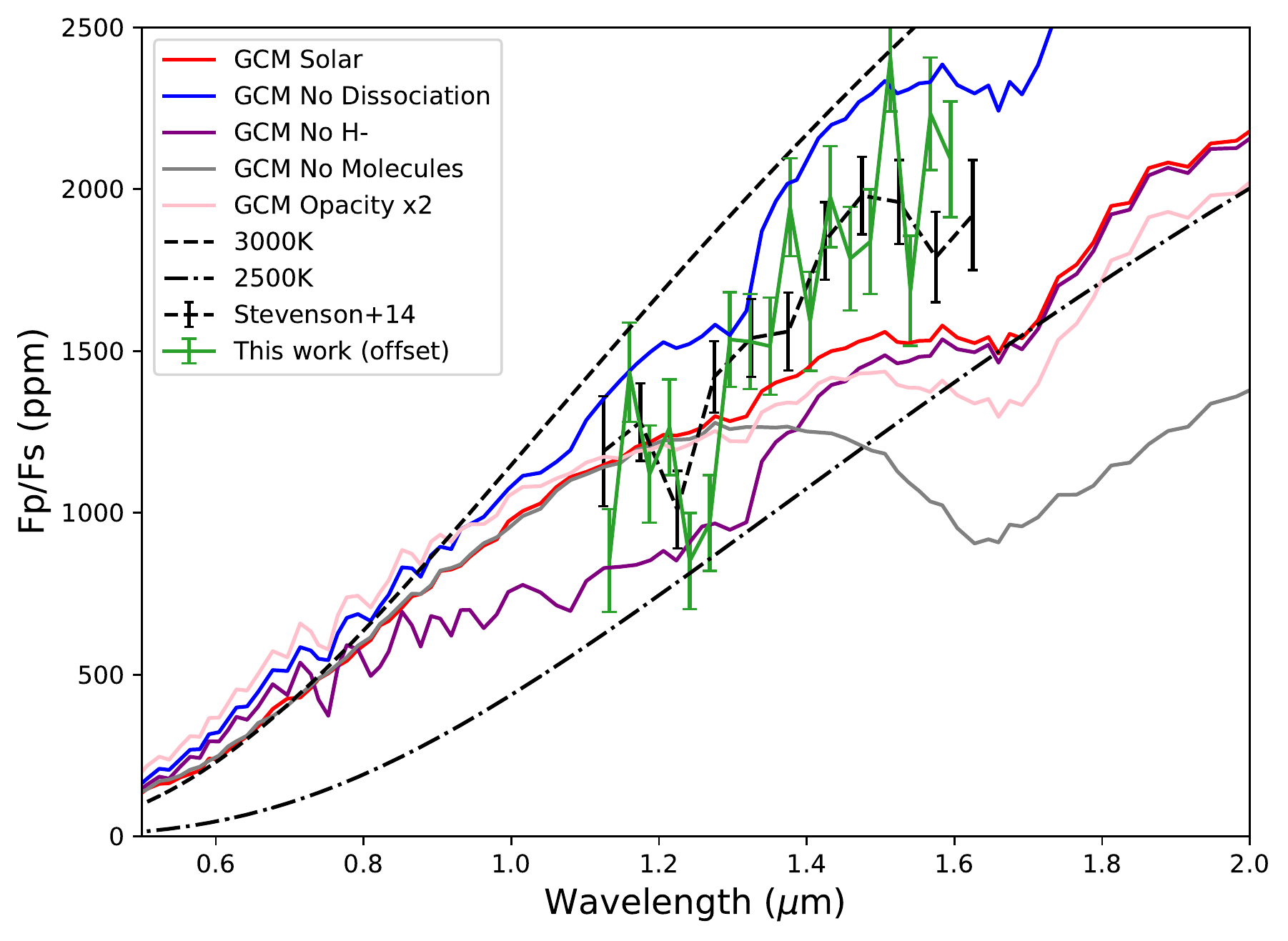}
\caption{Emission spectrum of WASP-12b as measured by HST/WFC3 (absolute measurement) compared to global circulation models and black-body emission curves. The Solar (red) and 2x  Optical Opacity (pink) models are self-consistent forward models. The remaining models include post-processing changes to the solar model. No dissociation (blue) has no dissociation of molecules, No H- (purple) has no H- opacities, No molecules (grey) has no molecular opacities and therefore mimics an H- continuum spectrum.}
\label{fig:abs_gcm}
\end{figure}

We also show the absolute spectra extracted by \citet{Stevenson2014b} compared to our circulation models also measured in absolute (Figure~\ref{fig:abs_gcm}). Here we can see that none of the models are able to match both the slope and the absolute brightness of the spectrum. 

\section{Discussion}
\label{Sec:Discussion}

\subsection{Emission spectra of WASP-12b compared to circulation models}
\label{Sec:slope}
We compare our data to a range of circulation models. Broadly, our data accurately match the model prediction of decreasing temperature from eclipse to quadrature due to the minimal day--night redistribution of flux. Our extracted quadrature spectrum provides a constraint on the temperature through the measured black-body temperature of 2124$\pm$417 K, but is not precise enough to distinguish between our generated models.

Our relative dayside spectrum has the slope of a black-body spectrum at 3186$\pm$677K, in agreement with the absolute emission spectrum obtained by \citet{Stevenson2014b}. The comparison of both extractions of the dayside spectrum to our nominal circulation model reveals a steeper spectrum than the models (Figure \ref{fig:eclipses}, \ref{fig:abs_gcm}). This slope seen in the data may be indicative of additional physics present in the climate of the planet that are not accounted for in our models, such as non-equilibrium chemistry. We briefly explore two possible physical effects that could be responsible for this increased slope, an increased temperature and non-equilibrium water abundances.

One explanation for the steeper slope seen in the dayside spectrum could be a genuinely hotter dayside temperature caused for instance by a steeper temperature and/or pressure profile than predicted by our models. Each of our models exhibits a temperature inversion due to a combination of optical absorbers and reduced cooling by thermal dissociation \citep{Lothringer2018, Parmentier2018} although a large temperature inversion on WASP-12b has not been detected \citep{Kreidberg2015}. Therefore, a steeper temperature--pressure profile could originate from an increase in optical opacity at low pressures, either due to a metallicity enhancement or non-equilibrium abundances. We produced an additional circulation model with enhanced optical opacities
(by a factor of two), mimicking an enhancement of metal compounds in the upper atmosphere (e.g. TiO), and forcing the temperature--pressure profile to be more strongly inverted. However, we see  that the increased temperature at low pressures causes a net loss in flux from the emission spectrum at wavelengths redder than 1.4$\mu$m (see the pink curve in Figure~\ref{fig:abs_gcm}). This is due to the increased dissociation of water at higher temperatures. This dissociation causes the observed water emission to be observed from deeper layers below the inversion where the atmosphere is cooler. It is therefore unlikely that an increase in the local temperature could cause the observed increased steepness of the emission spectrum, as the chemical response by dissociation to that increased temperature acts to reduce the slope of the spectrum. 

Another explanation could be non-equilibrium processes. Our models compute the dissociation fraction of water assuming equilibrium chemistry. However, the transport of cool gas from the nightside of the planet to the hot dayside could increase the contribution of water to our spectra if the dissociation of water is slower than its advection. Vertical transport of water from the deeper atmosphere is unlikely to play a dominant role, as the vertical winds are much weaker than the horizontal winds, and the temperature of the dayside is too hot for quenching to occur below $\sim$1 mbar \citep{Oreshenko2017}.
We estimate timescales for the thermal dissociation of water (as photo-dissociation is not expected at 0.1 bar levels) using the methods of \citet{Tsai2018}. To first order, the dissociation and recombination timescales are equal, because they are evaluated at equilibrium. We find that the rate-limiting reaction for water dissociation in this case is the thermal dissociation of H$_2$ to produce H atoms. This reaction produces H atoms, which allows water to dissociate through the very fast reaction of H + H$_2$O -> OH + H$_2$. The timescale for the thermal dissociation of hydrogen is therefore the same as the timescale for the dissociation of water at these temperatures. 

\begin{figure}
\includegraphics[scale=0.65]{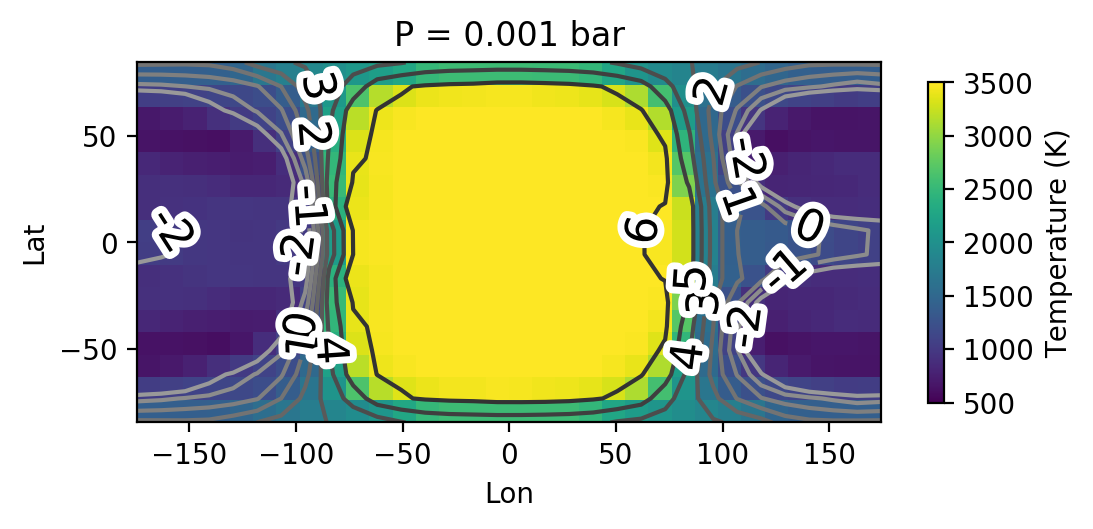}
\includegraphics[scale=0.65]{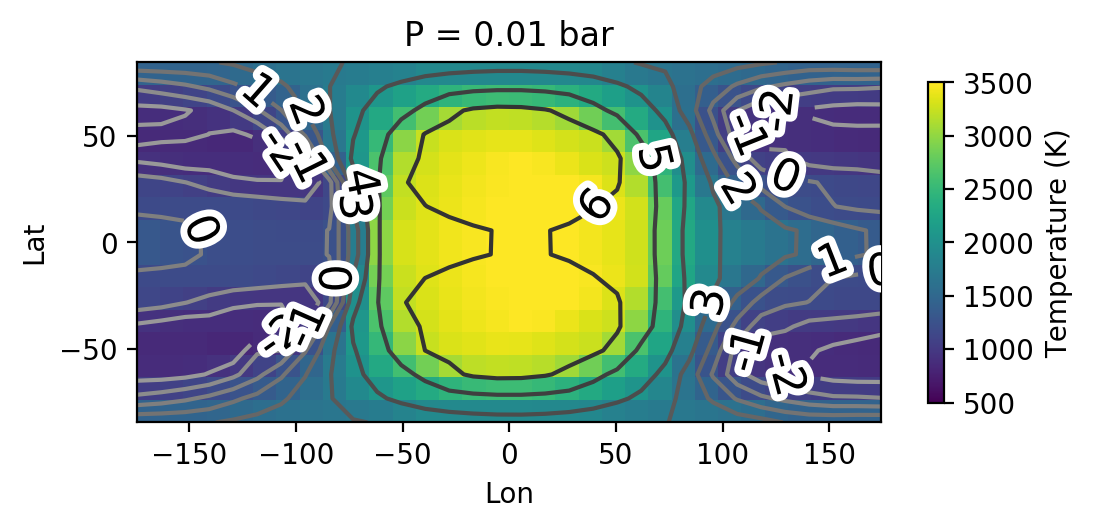}
\includegraphics[scale=0.65]{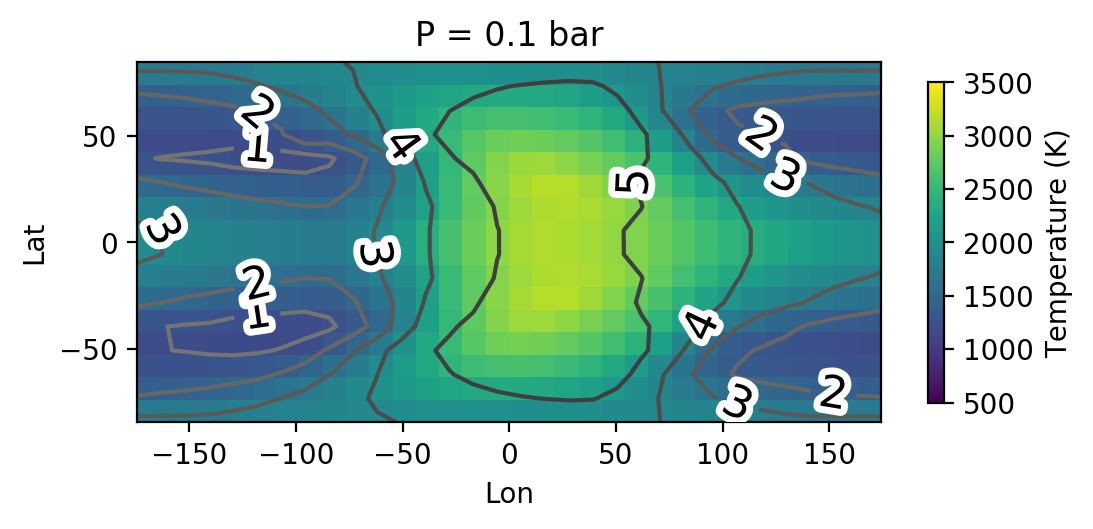}
\includegraphics[scale=0.65]{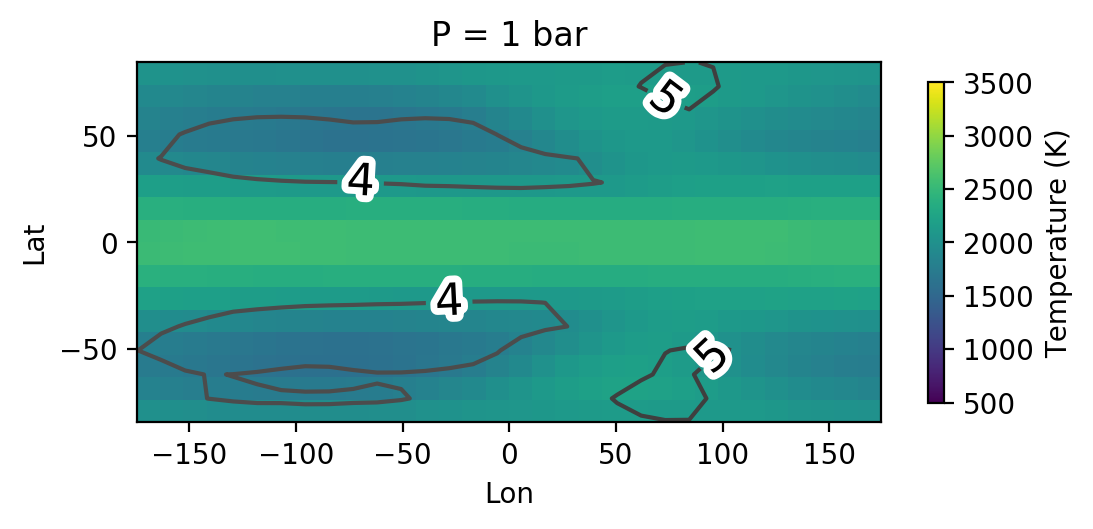}
\includegraphics[scale=0.65]{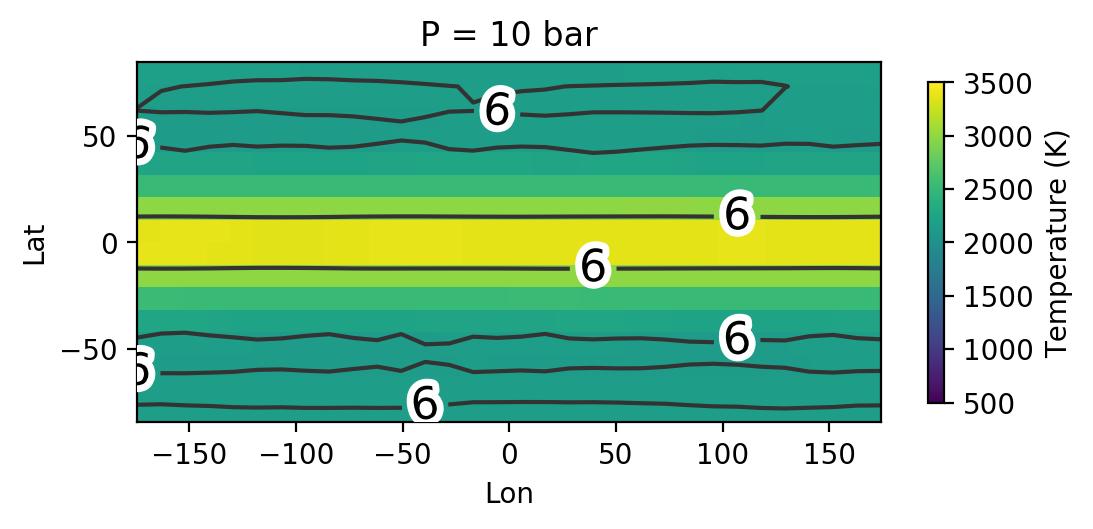}
\caption{Brightness temperature maps in 2D as produced by our GCM using solar abundances at five different pressures. Contours show the ratio of the simulated advective timescale over the predicted dissociation timescale for H$_2$, on a decimal log scale.}
\label{fig:timescales}
\end{figure}

We compare our estimated dissociation timescale to the advective timescale in our models. In this context the advective timescale is the ratio of the planet's radius to the magnitude of the UV components of the wind speed, as the vertical wind speeds are small. Figure~\ref{fig:timescales} shows contours of the logarithm of the ratio between the advective (wind) timescale and our calculated hydrogen dissociation timescales, plotted here over the background temperature map at three different pressures.
In this figure, it can be seen that the dayside advective timescale is always more than 100 times longer than the dissociation timescale, and typically between $10^3$ and $10^6$ times slower. Regions of the map with timescale ratios close to zero or negative can only be seen in the cooler regions of the atmosphere where dissociation will not take place. Therefore, any parcel of dissociated gas that could be carried to these regions to remain out of equilibrium would cross many contours, resulting in all of the dissociated gas recombining. We therefore conclude that water and H$_2$  likely reach their equilibrium dissociation fractions in the atmospheres of UHJs despite the winds of a few km/s. Globally, the dayside advective timescale is much longer than typical thermal dissociation timescales, suggesting that the dayside is in equilibrium with respect to other thermally dissociated molecules; however, this remains to be calculated for each species.

\subsection{Comparison to Spitzer phase curves of WASP-12b}

Previous studies of the phase curve of WASP-12b revealed a complicated system, as data show that the planet deviates from simple model predictions. Results from Spitzer photometry show a sinusoidal phase curve typical of an UHJ in the first bandpass around 3.6$\mu$m with a hot dayside and cool nightside, but two additional peaks at quadrature in the second bandpass at 4.5$\mu$m were found \citep{Cowan2012, Bell2019}. We compare the brightness temperatures of the planet's emission in Spitzer to our measured black-body temperatures derived from WFC3 data. We use the fit results from \citet{Bell2019} to determine the equivalent brightness at each HST orbit from both Spitzer bandpasses and convert this to a brightness temperature for the planet (Table~\ref{Tab:BrTps}). This allows a quantitative comparison of our work with previous studies. Differences between the results are expected because of the changes in opacities between the HST wavelength range and those measured by Spitzer, which lead to different pressure levels and species probed by the observations. We present this comparison below. 

The phase curves from Spitzer at 3.6 and 4.5 $\mu$m have been interpreted as evidence of gas stripped from the planet, which emits in the infrared and is responsible for the double peaked feature in the 4.5$\mu$m phase curve \citep{Cowan2012, Bell2019}.
This gas would need to be either sufficiently cool or CO rich for it to be visible in the 4.5$\mu$m phase curve but not at 3.6$\mu$m. Therefore, the natural prediction is that the double-peaked nature of the 4.5$\mu$m phase curve should not be seen in the HST/G141 bandpass, as at these wavelengths CO is not a dominant opacity, and any gas sufficiently cool to not be seen at 3.6 $\mu$m should also not affect the spectrum around 1.4 $\mu$m.
Our results are consistent with this hypothesis, as we measure a significant decrease in temperature going from the dayside to quadrature, in contrast to the double-peaked behaviour at 4.5$\mu$m.

While theoretically there is a CO band at 1.6$\mu$m, it is about 1000 times weaker than that at 4.5$\mu$m, resulting in a deviation of only a few ppm in the phase curve when extrapolating the Spitzer 4.5$\mu$m result. We are therefore not able to determine whether there is any additional CO emission from our spectra.

The 3.6$\mu$m phase curve exhibits different behaviour from that of the 4.5$\mu$m phase curve. At 3.6$\mu$m, the planet's emission is dominated by water, similar to the predictions for the HST WFC3/G141 bandpass. We therefore expect the behaviour and amplitude of the phase curve at 3.6$\mu$m to be similar to phase curves taken with HST WFC3/G141. The 3.6$\mu$m phase offset measured in Spitzer is seen to vary, but our HST results are more consistent with the observations from 2013 where only a small offset was seen. However, the planetary phases at which we measure temperatures in this work do not cover enough of the phase curve to measure a hot-spot offset, but remain consistent with a small westward offset, if any at all.

\bgroup

\renewcommand{\arraystretch}{1.2}
\begin{table*}[ht]
\begin{center}
\begin{tabular}{ | c | c | c | c | c | c |}
\hline
Instrument & & Dayside T$_p$ & & Quadrature T$_p$ & Observation \\
 & before & mean & after & K & Year \\
\hline 
& & & & & \vspace{-2ex} \\
HST G141 & 3108$\pm$520 & 3186$\pm$677 & 3270$\pm$557 & 2124$\pm$417 & 2011 \\
(this work) & & & & & \vspace{-0ex} \\
\hline 
Spitzer Ch1 & 2720$\pm$49 & 2522$\pm$44 & 2319$\pm$74 & 2340$\pm$75 & 2010 \\
(3.6$\mu$m) & 2400$\pm$70 & 2576$\pm$43 & 2746$\pm$51 & 2032$\pm$87 & 2013 \\
& & & & & \vspace{-2ex} \\
Spitzer Ch2 & 2759$\pm$59 & 2721$\pm$49 & 2684$\pm$80 & 2904$\pm$84 & 2010 \\
(4.5$\mu$m) & 2689$\pm$71 & 2612$\pm$47 & 2534$\pm$62 & 3024$\pm$91 & 2013 \\
\hline
\end{tabular}
\end{center}
\caption{Measured black-body temperatures from HST compared to brightness temperatures in Spitzer, binned to the same phases, derived from \citet{Bell2019}. Our method uses the slope of the emission spectrum to estimate the brightness temperature rather than the absolute emission, and does not marginalise over a given model, and therefore the uncertainities on the final temperatures are considerably larger.}
\label{Tab:BrTps}
\end{table*}
\egroup

\subsection{Pointing stability between visits}

We  further explore whether the pointing drift between visits can be explained and how this could be mitigated in future observations.

Our measured pointing offsets (shown in Figure~\ref{fig:shifts}) are within the nominal expectation for the target acquisition precision and fine guidance sensor (FGS) operation of the HST. The target acquisition was performed under the default mode for each visit, with a re-acquisition occurring after each earth occultation. This has an expected jitter of $\sim$7mas (~0.05 pix) as well as a $\sim$50mas drift over 12h or more (approximately 0.5 pixels) which is attributed to changes in the thermal properties of the FGS (from \citeauthor{Primer28}).

Hence, the measured drift within each orbit, as well as the offset between the first visit (quadrature) and the second visit (transit), are within the limits of nominal HST performance. Notably, the third visit (eclipse) is almost perfectly aligned in position with the first visit (quadrature). We first verified with STScI that no additional manoeuvres were performed and that FGS acquisition was executed as normal. We also found that the reference star HD258439, which was observed with two orbits at the end of each visit, exhibited the same movement on the detector as WASP-12 between visits, demonstrating that the FGS acquisition was not at fault. We therefore conclude that some change in the instrument, such as in the optical path through the location of the detector in the focal plane, occurred during the second visit alone and is responsible for the observed shift of the spectrum. We confirmed with STScI that such behaviour is within the expected performance of the telescope. This also implies that the very close alignment between the first and third visits was better than the nominal pointing precision of HST simply by chance.

 The likelihood that visits observed within a short time-span are well aligned on the detector is not clear from our observations.
However, it may be possible in the future to ensure that the position of the spectrum on the detector is well aligned with a previous visit, for example through routines such as a "real time interaction" target acquisition, which are not recommended due to the additional requirements on scheduling and man hours \citep{Primer28}.



\subsection{Partial phase curves in the future with HST, JWST, and other spectroscopic space facilities}

Full phase curves remain the best choice of method for obtaining longitudinally resolved absolute emission spectra of exoplanets in particular phase curves (i.e. those anchored by an eclipse at the start and end of the visit in order to allow for evaluation of any residual systematic
errors or drift). Currently, HST can only produce full phase curves for planets with periods of  $\sim$1
day, and JWST is only capable of this for planets with periods of less than 2 days, and so an alternative is needed, particularly for longer-period planets. Partial phase curves are a cheap alternative that may allow for some of the same science to be accomplished, in particular when looking for changes in spectral features with longitude.

Archival data of transits and eclipses will only be accessible with this technique should the chance alignment of the dispersed spectra on the detector be better than 0.1 pixels, assuming the visits were already taken with identical observing modes. Performance may be different for spatially scanned data, as there can be additional drift of the telescope within an exposure depending on the adopted scan rate of the telescope. The typical performance of the spatial scanning mode has been shown to result in drifts below $\sim$15mas ($\sim$0.11 pixels) for 77\% of data taken in this mode \citep{Stevenson2019}. Further study would be needed to verify that performance at this level, or with a lower pointing precision, would be sufficient to analyse spatially scanned data with this technique. However, it is  likely that at least 10\% of observations will not have a sufficient pointing stability for our method, as these observations are deemed to have failed, with a pointing drift of >1 pixel \citep{Stevenson2019}.

Both JWST and the ARIEL mission are scheduled to observe phase curves of exoplanets in the near future. As JWST observations are expected to operate at an increased pointing precision relative to HST of better than 17mas scatter per axis (JWST User Documentation), this technique should be applicable to JWST data. JWST will be in high demand, so full phase curves are only expected to be taken towards a few key targets. Additionally, these targets will be limited to short-period planets, as the single visit duration is limited to 48h. Therefore, common mode techniques such as ours will allow for some climate studies of planets for which full phase curves are not taken (e.g. see \citealt{Stevenson2020} for specific applications to JWST and non-transiting planets). Although the live performance of the telescope will not be known until data are analysed, in particular relating to any wavelength-dependent noise components, common-mode methods such as the technique presented here will be needed to characterise the performance alongside classical fits to the systematic
errors.

In summary partial phase curves will likely require dedicated observing programs to succeed, but should be considered for future spectroscopic space-based facilities, such as JWST and ARIEL. We recommend that special care be taken when observing targets in eclipse and transit in the future, as additional science beyond the goals of the program may be possible if their alignment on the detector is ensured.

\section{Conclusions}
\label{Sec:Conc}

We present the first emission spectrum of an exoplanet measured in quadrature outside of a phase curve. We measure the brightness as T=2124$\pm$417 K. 
We present our new technique for extracting planet spectra at different phases from spectroscopic phase curves based on common mode methods. We show how this technique succeeds in extracting spectra of WASP-12b obtained with HST/WFC3 G141 in emission at eclipse and quadrature, but also how it fails at transit where we find the data are offset in position on the detector. 

We obtain relative emission spectra of WASP-12b at eclipse and quadrature, and find that the planet exhibits a decrease in temperature with longitude expected for nominal day--night redistribution of 1062$\pm$516 K. This result is in line with the findings from full-orbit Spitzer phase curves, which suggest the presence of cool gas in the system that is not visible at shorter wavelengths \citep{Bell2019}. 

We find that our extracted spectrum is steeper than expected from global circulation models, and while we do not have a clear explanation for this, we find that this cannot be explained by non-equilibrium water fractions or increased local temperature. We also calculate from our global circulation models that the dayside atmospheric pressures probed by our observations should be in equilibrium with respect to thermal dissociation of water and hydrogen. 

We outline how to achieve success with this data analysis technique in the future, either through continuous phase curve observations or by careful monitoring of the PSF of individual visits in a partial phase curve. This will likely be a useful observing method for JWST and ARIEL, where partial phase curves have a greater chance of succeeding. Our new technique may also create an opportunity for further studies on archival data, as well verification of existing results from full-orbit phase curves.

\section*{Acknowledgements}
We would like to thank the STScI helpdesk and in particular Varun Bajaj for fielding our many questions on the WFC3 performance. We would also like the thank Dr. Channon Visscher for their insights on dissociation timescales.

J.M.D acknowledges support from the Amsterdam Academic Alliance (AAA) Program, and the European Research Council (ERC) European Union’s Horizon 2020 research and innovation programme (grant agreement no. 679633; Exo-Atmos). This work is part of the research programme VIDI New Frontiers in Exoplanetary Climatology with project number 614.001.601, which is (partly) financed by the Dutch Research Council (NWO).

\section*{Appendix}

\begin{table}[ht]
\begin{center}
\begin{tabular}{ | c | c  c | c  c |}
\hline
 Central & Eclipse & (ppm) & Quad & (ppm) \\
 $\lambda$ & Fp/Fs & Error & Fp/Fs & Error \\

 \hline
1.14 & -750 & 159 & -553 & 224 \\
1.16 & -34 & 152 & 38 & 216 \\
1.19 & -602 & 150 & -62 & 212 \\
1.22 & -390 & 148 & 89 & 210 \\
1.25 & -617 & 149 & -506 & 211 \\
1.27 & -590 & 147 & 285 & 208 \\
1.30 & -69 & 147 & 70 & 207 \\
1.33 & -69 & 147 & -577 & 208 \\
1.35 & 14 & 150 & -204 & 213 \\
1.38 & 269 & 152 & 126 & 215 \\
1.41 & 131 & 154 & -2 & 217 \\
1.44 & 287 & 157 & 491 & 222 \\
1.46 & 276 & 161 & -85 & 227 \\
1.49 & 143 & 162 & -14 & 229 \\
1.52 & 891 & 167 & 242 & 235 \\
1.55 & 76 & 171 & -157 & 241 \\
1.57 & 700 & 174 & 543 & 246 \\
1.60 & 332 & 180 & 274 & 255 \\
\hline
\end{tabular}
\end{center}
\caption{Extracted eclipse and quadrature spectra.}
\label{Tab:Specs}
\end{table}

\begin{table}[ht]
\begin{center}
\begin{tabular}{ | c | c | c |}
\hline
Wavelength & Dilution (\%) & Error \\
\hline
1.133 & 3.39 & 0.13 \\
1.160 & 4.68 & 0.13 \\
1.187 & 4.81 & 0.13 \\
1.214 & 5.04 & 0.14 \\
1.242 & 5.39 & 0.14 \\
1.269 & 5.71 & 0.15 \\
1.296 & 5.89 & 0.15 \\
1.323 & 5.97 & 0.15 \\
1.350 & 6.14 & 0.17 \\
1.377 & 5.55 & 0.18 \\
1.405 & 5.74 & 0.20 \\
1.432 & 5.39 & 0.21 \\
1.459 & 5.51 & 0.23 \\
1.486 & 5.74 & 0.24 \\
1.513 & 5.70 & 0.23 \\
1.541 & 6.09 & 0.26 \\
1.568 & 6.47 & 0.26 \\
1.595 & 6.77 & 0.28 \\
\hline
\end{tabular}
\end{center}
\caption{Dilution correction for the contribution of WASP-12BC to the planet spectra. The dilution here is defined as the flux of the companions relative to the flux of WASP-12A, shown here as a percentage.}
\label{Tab:Dilution}
\end{table}


\begin{thebibliography}{}

\bibitem[Arcangeli et al.(2018)]{Arcangeli2018} Arcangeli, J., D{\'e}sert, J.-M., Line, M.~R., et al.\ 2018, \apjl, 855, L30 
\bibitem[Arcangeli et al.(2019)]{Arcangeli2019} Arcangeli, J., D{\'e}sert, J.-M., Parmentier, V., et al.\ 2019, \aap, 625, A136
\bibitem[Baxter et al.(2020)]{Baxter2020} Baxter, C., D{\'e}sert, J.-M., Parmentier, V., et al.\ 2020, \aap, 639, A36
\bibitem[Bell et al.(2017)]{Bell2017} Bell, T.~J., Nikolov, N., Cowan, N.~B., et al.\ 2017, \apjl, 847, L2 
\bibitem[Bell et al.(2019)]{Bell2019} Bell, T.~J., Zhang, M., Cubillos, P.~E., et al.\ 2019, \mnras, 489, 1995
\bibitem[Borucki et al.(2009)]{Borucki2009} Borucki, W.~J., Koch, D., Jenkins, J., et al.\ 2009, Science, 325, 709 
\bibitem[Bergfors et al.(2011)]{Bergfors2011} Bergfors, C., Brandner, W., Henning, T., et al.\ 2011, The Astrophysics of Planetary Systems: Formation, Structure, and Dynamical Evolution, 397
\bibitem[Bergfors et al.(2013)]{Bergfors2013} Bergfors, C., Brandner, W., Daemgen, S., et al.\ 2013, \mnras, 428, 182
\bibitem[Bushouse(2008)]{Bushouse2008} Bushouse, H.\ 2008, WFC3 Instrument Science Report 2008-28, 14 pages
\bibitem[Charbonneau et al.(2005)]{Charbonneau2005} Charbonneau, D., Allen, L.~E., Megeath, S.~T., et al.\ 2005, \apj, 626, 523
\bibitem[Cowan et al.(2012)]{Cowan2012} Cowan, N.~B., Machalek, P., Croll, B., et al.\ 2012, \apj, 747, 82
\bibitem[Crossfield et al.(2012)]{Crossfield2012a} Crossfield, I.~J.~M., Hansen, B.~M.~S., \& Barman, T.\ 2012, \apj, 746, 46
\bibitem[Deming et al.(2006)]{Deming2006} Deming, D., Harrington, J., Seager, S., et al.\ 2006, \apj, 644, 560
\bibitem[Deming et al.(2013)]{Deming2013} Deming, D., Wilkins, A., McCullough, P., et al.\ 2013, \apj, 774, 95 
\bibitem[Mikal-Evans et al.(2019)]{Evans2019} Mikal-Evans, T., Sing, D.~K., Goyal, J.~M., et al.\ 2019, \mnras, 488, 2222

\bibitem[HST Primer for Cycle 28 (2020)]{Primer28} Space Telescope Science Institute 2020, HST Primer for Cycle 28 
\bibitem[Kitzmann et al.(2018)]{Kitzmann2018} Kitzmann, D., Heng, K., Rimmer, P.~B., et al.\ 2018, \apj, 863, 183
\bibitem[Knutson et al.(2007)]{Knutson2007} Knutson, H.~A., Charbonneau, D., Allen, L.~E., et al.\ 2007, \nat, 447, 183 
\bibitem[Kreidberg et al.(2015)]{Kreidberg2015} Kreidberg, L., Line, M.~R., Bean, J.~L., et al.\ 2015, \apj, 814, 66 
\bibitem[Kreidberg et al.(2018)]{Kreidberg2018} Kreidberg, L., Line, M.~R., Parmentier, V., et al.\ 2018, arXiv:1805.00029 
\bibitem[Kurucz(1993)]{Kurucz1993} Kurucz, R.~L.\ 1993, VizieR Online Data Catalog, VI/39
\bibitem[Lothringer et al.(2018)]{Lothringer2018} Lothringer, J.~D., Barman, T., \& Koskinen, T.\ 2018, \apj, 866, 27.
\bibitem[Mandell et al.(2013)]{Mandell2013} Mandell, A.~M., Haynes, K., Sinukoff, E., et al.\ 2013, \apj, 779, 128
\bibitem[Mansfield et al.(2018)]{Mansfield2018} Mansfield, M., Bean, J.~L., Line, M.~R., et al.\ 2018, \aj, 156, 10 
\bibitem[Oreshenko et al.(2017)]{Oreshenko2017} Oreshenko, M., Lavie, B., Grimm, S.~L., et al.\ 2017, \apjl, 847, L3
\bibitem[Parmentier et al.(2018)]{Parmentier2018} Parmentier, V., Line, M.~R., Bean, J.~L., et al.\ 2018, \aap, 617, A110 
\bibitem[STScI Development Team(2013)]{pysynphot} STScI Development Team\ 2013, Astrophysics Source Code Library

\bibitem[Showman et al.(2009)]{Showman2009} Showman, A.~P., Fortney, J.~J., Lian, Y., et al.\ 2009, \apj, 699, 564 
\bibitem[Snellen et al.(2009)]{Snellen2009} Snellen, I.~A.~G., de Mooij, E.~J.~W., \& Albrecht, S.\ 2009, \nat, 459, 543 
\bibitem[Stevenson et al.(2014a)]{Stevenson2014a} Stevenson, K.~B., Bean, J.~L., Seifahrt, A., et al.\ 2014, \aj, 147, 161 
\bibitem[Stevenson et al.(2014b)]{Stevenson2014b} Stevenson, K.~B., Bean, J.~L., Madhusudhan, N., et al.\ 2014, \apj, 791, 36 
\bibitem[Stevenson et al.(2014c)]{Stevenson2014c} Stevenson, K.b., D{\'e}sert, J.-M., Line, M.~R., et al.\ 2014, Science, 346, 838 
\bibitem[Stevenson et al.(2017)]{Stevenson2017} Stevenson, K.~B., Line, M.~R., Bean, J.~L., et al.\ 2017, \aj, 153, 68
\bibitem[Stevenson, \& Fowler(2019)]{Stevenson2019} Stevenson, K.~B., \& Fowler, J.\ 2019, Space Telescope WFC Instrument Science Report
\bibitem[Stevenson (2020)]{Stevenson2020} Stevenson, K.~B. \& Space Telescopes Advanced Research Group on the Atmospheres of Transiting Exoplanets\ 2020, \apjl, 898, L35. doi:10.3847/2041-8213/aba68c
\bibitem[Swain et al.(2013)]{Swain2013} Swain, M., Deroo, P., Tinetti, G., et al.\ 2013, \icarus, 225, 432 
\bibitem[Tsai et al.(2018)]{Tsai2018} Tsai, S.-M., Kitzmann, D., Lyons, J.~R., et al.\ 2018, \apj, 862, 31
\bibitem[Tsiaras et al.(2016)]{Tsiaras2016} Tsiaras, A., Waldmann, I.~P., Rocchetto, M., et al.\ 2016, \apj, 832, 202
\bibitem[Tsiaras \& Ozden(2019)]{Tsiaras2019} Tsiaras, A., \& Ozden, J.\ 2019, arXiv e-prints, arXiv:1908.01692
\bibitem[Wakeford et al.(2016)]{Wakeford2016} Wakeford, H.~R., Sing, D.~K., Evans, T., et al.\ 2016, \apj, 819, 10 
\bibitem[Zhou et al.(2017)]{Zhou2017} Zhou, Y., Apai, D., Lew, B.~W.~P., et al.\ 2017, \aj, 153, 243 


\end{thebibliography}
\end{document}